# The Gaussian Interference Relay Channel: Improved Achievable Rates and Sum Rate Upperbounds Using a Potent Relay

Ye Tian and Aylin Yener
Wireless Communications and Networking Laboratory
Electrical Engineering Department
The Pennsylvania State University, University Park, PA 16802
*yetian@psu.edu* *yener@ee.psu.edu*

**Abstract**

We consider the Gaussian interference channel with an intermediate relay as a main building block for cooperative interference networks. On the achievability side, we consider compress-and-forward based strategies. Specifically, a generalized compress-and-forward strategy, where the destinations jointly decode the compression indices and the source messages, is shown to improve upon the compress-and-forward strategy which sequentially decodes the compression indices and source messages, and the recently proposed generalized hash-and-forward strategy. We also construct a nested lattice code based compute-and-forward relaying scheme, which outperforms other relaying schemes when the direct link is weak. In this case, it is shown that, with a relay, the interference link can be useful for decoding the source messages. Noting the need for upperbounding the capacity for this channel, we propose a new technique with which the sum rate can be bounded. In particular, the sum capacity is upperbounded by considering the channel when the relay node has abundant power and is named potent for that reason. For the Gaussian interference relay channel with potent relay, we study the strong and the weak interference regimes and establish the sum capacity, which, in turn, serve as upperbounds for the sum capacity of the GIFRC with finite relay power. Numerical results demonstrate that upperbounds are tighter than the cut-set bound, and coincide with known achievable sum rates for many scenarios of interest. Additionally, the degrees of freedom of the GIFRC are shown to be 2 when the relay has large power, achievable using compress-and-forward.

This work was supported by the National Science Foundation with grants CCF 0237727, CNS 0716325, CNS 0721445, CNS 0964364 and the DARPA ITMANET Program with Grant W911NF-07-1-0028. This work was presented in part at IEEE Globecom 2009 [1], and IEEE ICC 2010 [2].

## I. Introduction

Wireless medium allows signals transmitted from one user to be overheard by surrounding users. This fact causes interference between different user pairs, but can also be utilized to facilitate cooperation between the nodes. The interference relay channel (IFRC) is a fundamental model that addresses the case when interference and cooperation co-exist in the same network.

The IFRC consists of two senders with two corresponding receivers, and an intermediate relay. Reference [3] has proposed an achievable scheme based on rate splitting at the sources and letting the relay decode both the common and the private messages to help both sources. A modified model is also proposed in [4] and [5], where the relay is cognitive to the messages at both sources. In [4], beamforming, dirty paper coding and time sharing are used to obtain the achievable region. In [5], Han-Kobayashi coding and dirty paper coding are combined to improve the rate in [4]. Another achievability technique is interference forwarding developed in [6] and [7], which demonstrates that forwarding interference can be beneficial. The capacity results up to date are for special cases of IFRC [6], and the capacity region in general is open, as the channel seems to inherit the challenges of both the interference channel [8], [9] and the relay channel [10]. In this work, we make progress in characterizing the capacity of Gaussian IFRC (GIFRC): we provide improved achievable rates, and a sum capacity upperbound that is non-trivial, i.e., tighter than the cut set bound.

First, we focus on achievability. The limitation of DF relaying [3], [6], [7], [11] is that its performance is limited by the decoding capability of the relay. As a result, when the SNR of the received signal at the relay is low, the rates that can be achieved are small. To overcome this, we first consider using the compress-and-forward (CF) strategy in [10] with rate splitting at the sources, in order to mitigate the effect of interference. We obtain insights regarding how to treat interference with CF based relaying. We next note that for the IFRC, the destinations may have different side information, and the performance of the CF strategy that requires both destinations to recover the unique compression index is limited by the destination with the worst side information. This is also shown in recent reference [12]. To address this issue, we propose a generalized compress-and-forward (GCF) strategy for the IFRC, which generalizes the one used for relay channel in [13]. The GCF scheme also uses Wyner-Ziv coding, but does not need to use the side information at the destinations to uniquely decode any compression indices. Although this approach is shown to achieve the same rate as the CF strategy in [10] for the relay channel (see [13]), it achieves a larger rate region than the CF strategy in [10] for the IFRC. We also compare the GCF strategy with two strategies from two recent references [12], [14]. We show that the GCF scheme outperforms the generalized hash-and-forward in [12]. The GCF has lower coding complexity than and performs very close to noisy network coding in [14].

We next observe that, when the direct links are weak, Wyner-Ziv coding based CF relaying strategies have less than desirable performance. For this case, we design a nested lattice code based compute-and-forward relaying strategy. Specifically, we show that the interference links are useful for decoding the source messages. We show that this strategy can achieve higher sum rates than both DF and CF based relaying strategies and noisy network coding.

To measure how close we are to capacity with the proposed achievable rates, we need to upper bound the capacity as well. To do so, we next advocate for a GIFRC model that allows us to derive new sum capacity upper bounds. In particular, we consider the case where the relay node has very large (infinite) power. We term such a node a *potent relay*. In practice, the GIFRC with potent relay can be thought of a system where the relay node is a base station or access point that aids an ad hoc network, and its power constraint is much larger compared to those of the other transmitters. From the information theoretic perspective, the (sum) capacity of GIFRC with infinite relay power is clearly an upperbound for the one with finite relay power, and we show, in this paper, that it is a useful one. We first observe the equivalence between this model and the GIFRC with in-band reception/out-of-band noiseless transmission, with respect to the classification in [15]. To establish the sum capacity of this channel, we consider strong and weak interference regimes. To bound the sum rate in weak interference, we utilize a genie aided approach where the information is carefully optimized to yield a “smart and useful genie” [16]. For strong interference, we again use a genie argument [8]. For both cases, we show that the upperbounds are achievable, thus establishing the sum capacity for GIFRC with potent relay. Both results, in turn, serve as upperbounds for the sum capacity of the general GIFRC. Although our tight results are for when the relay has infinite power, we demonstrate that with finite relay power, the sum-rate upperbounds are tighter than the cut-set bound and numerically coincides with the achievable rates for many scenarios of interest. We also observe that the degrees of freedom of GIFRC increase from $1$ to $2$ when the power of the relay $P_R$ and the power of the sources $P$ satisfy $P_R = \mathcal{O}(P^2)$, or $P_R = 2P$ in dB, achievable using CF based relaying.

The remainder of the paper is organized as follows: Section II describes the channel model. Section III describes the achievable schemes based on CF. Section IV describes the compute-and-forward based achievable scheme. Section V introduces the notion of *potent relay* and shows the equivalence between GIFRC with potent relay and GIFRC with in-band reception/out-of-band noiseless transmission. Section VI establishes the sum capacity of GIFRC with potent relay in weak interference. Section VII establishes the sum capacity of GIFRC with potent relay in strong interference. Section X compares our potent relay based upperbounds with cut-set bound and various achievable schemes for GIFRC. Section XI presents concluding remarks.

## II. System model

### A. The Discrete Memoryless Interference Relay Channel (DM-IFRC)

First we describe the discrete memoryless interference relay channel (DM-IFRC), since the CF based relaying scheme is derived for this model and then specialized to the Gaussian case. We have three finite input alphabets $\mathcal{X}_1, \mathcal{X}_2, \mathcal{X}_R$, three output alphabets $\mathcal{Y}_1, \mathcal{Y}_2, \mathcal{Y}_R$, and a probability distribution

$$p(y_1^n, y_2^n, y_R^n | x_1^n, x_2^n, x_R^n) = \prod_{i=1}^{n} p(y_{1,i}, y_{2,i}, y_{R,i} | x_{1,i}, x_{2,i}, x_{R,i}) \tag{1}$$

which characterizes the channel. Each source $S_i$, $i = 1, 2$ wishes to communicate with its paired destination $D_j$, $j = 1, 2$. $S_i$ chooses a message $W_i$ from a message set $\mathcal{W}_i = \{1, 2, \ldots, 2^{nR_i}\}$, encodes this message into a length $n$ codeword with an encoding function $f_i(W_i) = X_i^n$ and transmits the codeword through the channel. The relay employs an encoding function based on the information it received from previous transmissions, i.e., $X_{R,t} = f_R(Y_R^{t-1})$. Each destination uses a decoding function $g_i(Y_i^n) = \hat{W}_i$. A rate pair $(R_1, R_2)$ is called achievable if there exists a message set, equipped with a set of encoding and decoding functions described above such that the error probability $\Pr(\hat{W}_1 \neq W_1 \bigcup \hat{W}_2 \neq W_2) \to 0$ as $n \to \infty$.

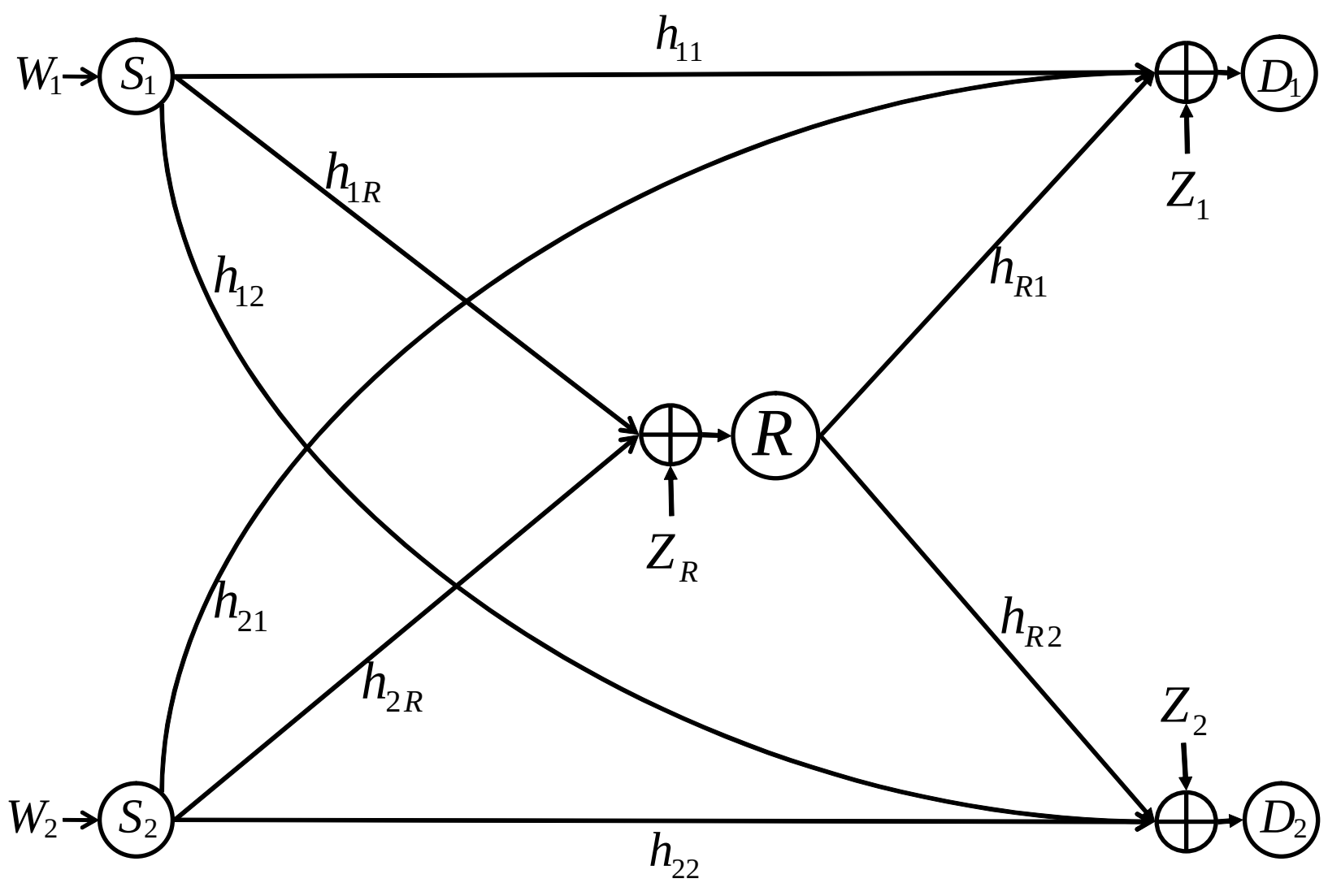

Fig. 1. The Gaussian Interference Relay Channel (GIFRC).

### *B. The Gaussian Interference Relay Channel (GIFRC)*

The Gaussian IFRC is shown in Fig. 1. The received signals at both destinations and the relay are:

$$Y_1 = h_{11}X_1 + h_{21}X_2 + h_{R1}X_R + Z_1 \tag{2}$$

$$Y_2 = h_{12}X_1 + h_{22}X_2 + h_{R2}X_R + Z_2 \tag{3}$$

$$Y_R = h_{1R}X_1 + h_{2R}X_2 + Z_R \tag{4}$$

Here, $Z_i$ $(i = 1, 2, R)$ are zero-mean, unit variance, independent Gaussian random variables that model the additive noise at each receiver. The channel gains are positive real numbers. The power constraints for the sources and the relay are $\frac{1}{n}\sum_{k=1}^{n} X_{i,k}^2 \leq P_i$, $i = 1, 2$, and $\frac{1}{n}\sum_{k=1}^{n} X_{R,k}^2 \leq P_R$, respectively.

## III. Achievable Strategies and Rate Regions

In this section, we first present a strategy which uses CF relaying in [10] and rate splitting at the sources [9]. We obtain insights from this case on how to treat interference for different channel parameters. We then focus on the impact of different relaying strategies on the achievable rates. We propose a generalized compress-and-forward (GCF) relaying scheme which improves upon the CF strategy. We also compare the rate regions obtained from GCF relaying and the generalized hash-and-forward (GHF) strategy in [12] and noisy network coding (NNC) in [14].

### A. Compress-and-Forward Relaying with Rate Splitting

The following rate region is based on the CF strategy in [10] with rate splitting at the sources to mitigate interference.

*Theorem 1:* The following rate tuples are achievable for the general interference relay channel with $R_1 = R_{10} + R_{11}$, $R_2 = R_{20} + R_{22}$:

$$R_{ii} \leq I(X_i; \hat{Y}_R, Y_i | U_i, U_j, X_R) \tag{5}$$

$$R_{i0} + R_{ii} \leq I(X_i; \hat{Y}_R, Y_i | U_j, X_R) \tag{6}$$

$$R_{ii} + R_{j0} \leq I(U_j, X_i; \hat{Y}_R, Y_i | U_i, X_R) \tag{7}$$

$$R_{i0} + R_{ii} + R_{j0} \leq I(U_j, X_i; \hat{Y}_R, Y_i | X_R) \tag{8}$$

where $i, j \in \{1, 2\}$, $i \neq j$, subject to

$$\min\{I(X_R; Y_1), I(X_R; Y_2)\} \geq \max\{I(Y_R; \hat{Y}_R | X_R, Y_1), I(Y_R; \hat{Y}_R | X_R, Y_2)\} \tag{9}$$

for all joint probability distributions

$$p(u_1)p(u_2)p(x_1|u_1)p(x_2|u_2)p(x_R)p(y_1 y_2 y_R | x_1 x_2 x_R)p(\hat{y}_R | y_R x_R)$$

*Proof:* See Appendix A. ∎

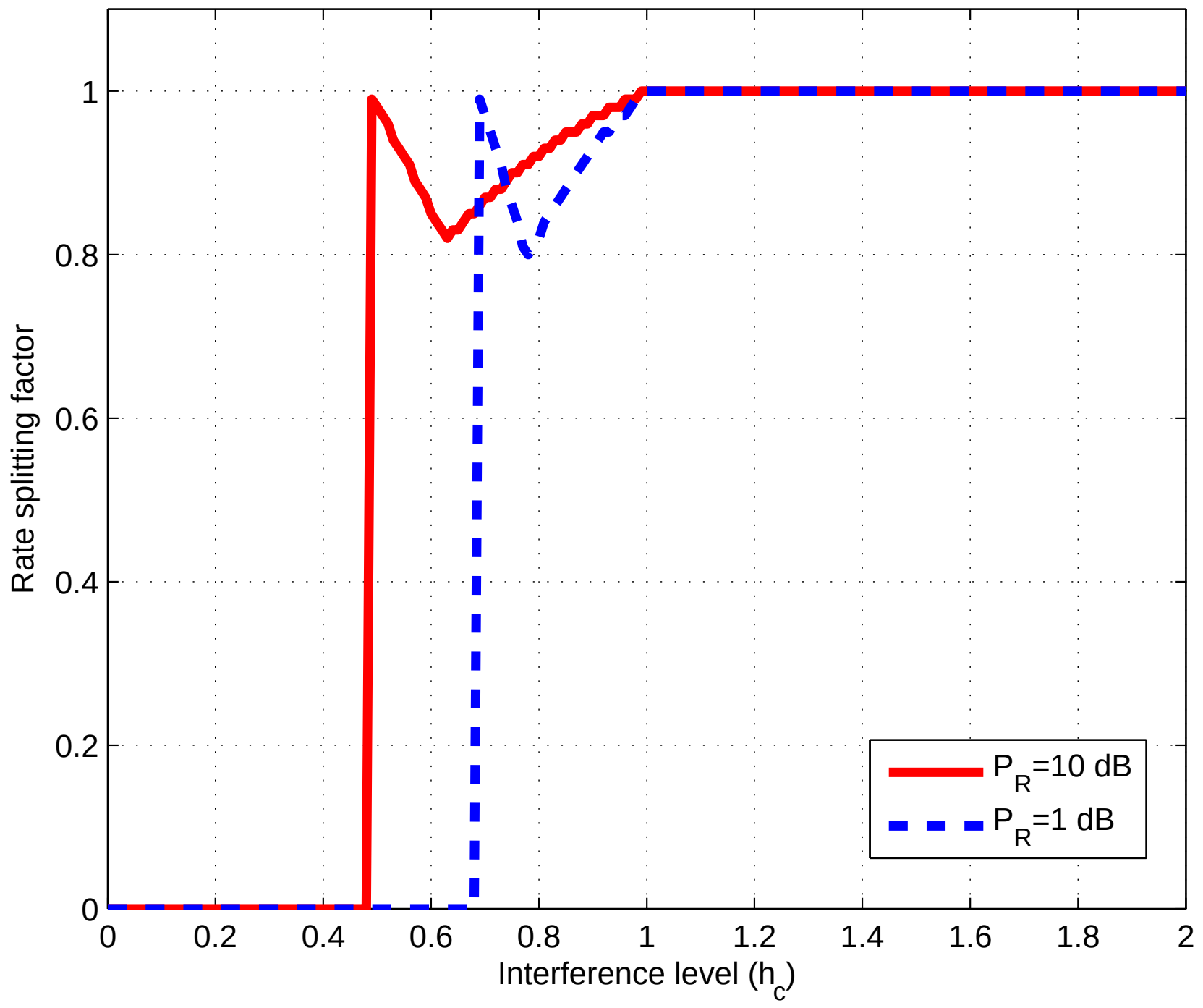

Fig. 2. Optimal rate splitting factor under different channel conditions when $h_{11} = h_{22} = h_d = 1$, $h_{12} = h_{21} = h_c$, $h_{1R} = h_{2R} = h_s = 1$, $h_{R1} = h_{R2} = h_R = 1$, $P = 1$dB.

*Remark 1:* The result can be extended to the Gaussian case by setting $U_1 \sim \mathcal{N}(0, \alpha P_1)$, $U_2 \sim \mathcal{N}(0, \beta P_2)$, $V_1 \sim \mathcal{N}(0, (1-\alpha)P_1)$, $V_2 \sim \mathcal{N}(0, (1-\beta)P_2)$, $X_R \sim \mathcal{N}(0, P_R)$, which are all independent from each other, and $X_1 = U_1 + V_1$, $X_2 = U_2 + V_2$, $\hat{Y}_R = Y_R + \hat{Z}_R$, where $\hat{Z}_R \sim \mathcal{N}(0, \sigma_R^2)$. Here, $U_1, U_2$ represent the common messages to be decoded at both receivers, whereas $V_1, V_2$ represent the private messages to be decoded at the intended receivers only. $\hat{Y}_R$ represents the compressed version of the received signal at the relay, which is to be forwarded to the receivers.

To gain further insight regarding how to treat the interference, we plot the optimal rate splitting factor versus the interference level in Figure 2 in the symmetric setting. Relay-destination ($R-D$) gains and transmit powers are set to unity. We can see that the way we treat interference is related to the interference level. When the interference link gain is weak and below a threshold, the optimal power allocation dictates to use all the power to transmit the private messages. This threshold depends on the power constraint of the relay. By contrast, when the interference is strong and above a threshold, the optimal power allocation dictates to use all the power to transmit the common messages, similar to the compound MAC in the 2-user strong interference channel [8]. We note that, for the symmetric case, a sufficient condition for strong interference is $h_c \geq h_d$. This can be verified by calculating the optimum factor $\alpha$ under the assumption $h_c \geq h_d$. This observation means that when $h_c \geq h_d$, we should allocate all the power to transmit the common messages regardless of the values of the remaining channel gains. For the asymmetric setting, the condition for strong interference depends also on the quality of source-relay and relay-destination links.

### *B. Generalized Compress-and-Forward Relaying*

The performance of the CF scheme requires the destinations to uniquely recover the compression index. As such, its performance is limited by the destination with the worst side information. To address this issue, we propose a generalized compress-and-forward scheme, where the destinations do no need to decode any compression indices. Specifically, after the destinations decode the bin index, we can use all the sequences in this bin to decode the source messages. In [13], this approach is shown to achieve the same rate as the CF in [10], but as we show later, this approach can achieve larger rate region for the IFRC, since the destinations have different side information. We can obtain the following achievable rate regions. Note that to focus on the advantage of the relaying strategy and for the clarity of exposition, we do not use rate splitting for the GCF strategy. Instead, we derive two achievable rate regions by either treating interference as noise, or trying to decode the interference.

*Theorem 2:* The following two rate regions are achievable using GCF.

$GCF_1$*: Destinations treat interference as noise*

$$R_i < \min\left\{I(X_i;\hat{Y}_R Y_i|X_R), I(X_i;Y_i|X_R) + R_0 - I(Y_R;\hat{Y}_R|X_i X_R Y_i)\right\} \tag{10}$$

where $i \in \{1,2\}$,

$$R_0 = \min\{I(X_R;Y_1), I(X_R;Y_2)\} \tag{11}$$

for all distributions

$$p(x_1)p(x_2)p(x_R)p(\hat{y}_R|y_Rx_R)p(y_1y_2y_R|x_1x_2x_R) \tag{12}$$

*$GCF_2$: Destinations try to decode the interference*

$$R_i < \min\left\{I(X_i;\hat{Y}_RY_i|X_jX_R), I(X_i;Y_i|X_jX_R) + R_0 - I(Y_R;\hat{Y}_R|X_1X_2X_RY_i)\right\} \tag{13}$$

$$R_1 + R_2 < \min\left\{I(X_1X_2;\hat{Y}_RY_i|X_R), I(X_1X_2;Y_i|X_R) + R_0 - I(Y_R;\hat{Y}_R|X_1X_2X_RY_i)\right\} \tag{14}$$

where $i,j \in \{1,2\}$, $i \neq j$,

$$R_0 = \min\{I(X_R;Y_1), I(X_R;Y_2)\} \tag{15}$$

for all distributions

$$p(x_1)p(x_2)p(x_R)p(\hat{y}_R|y_Rx_R)p(y_1y_2y_R|x_1x_2x_R) \tag{16}$$

*Proof:* See Appendix B. ■

In a concurrent work [12], the authors proposed a generalized hash-and-forward (GHF) relaying scheme. The scheme uses Wyner-Ziv coding as in the CF scheme, and each destination decodes a list of compression indices. It is shown that GHF has better performance than the CF scheme in high SNR regime, but the CF scheme has better performance in low SNR. Reference [12] considered the achievable strategy by treating interference as noise. When destinations try to decode interference, we can obtain the following rate region using GHF.

$$R_i < I(X_i;Y_i|X_jX_R) + R_0 - I(Y_R;\hat{Y}_R|X_1X_2X_RY_i) \tag{17}$$

$$R_1 + R_2 < I(X_1X_2;Y_i|X_R) + R_0 - I(Y_R;\hat{Y}_R|X_1X_2X_RY_i) \tag{18}$$

where $i,j \in \{1,2\}$, $i \neq j$,

$$R_0 = \min\{I(X_R;Y_1), I(X_R;Y_2)\} \leq \min\{I(Y_R;\hat{Y}_R|X_RY_1), I(Y_R;\hat{Y}_R|X_RY_2)\} \tag{19}$$

for all distributions

$$p(x_1)p(x_2)p(x_R)p(\hat{y}_R|y_Rx_R)p(y_1y_2y_R|x_1x_2x_R) \tag{20}$$

The rate region due to treating interference as noise has similar form [12, Equations (19)-(20)], with two differences: there is no sum rate bound and (17) is replaced with

$$R_i < I(X_i; Y_i|X_R) + R_0 - I(Y_R; \hat{Y}_R|X_i X_R Y_i), \tag{21}$$

It can be readily verified that when destinations treat interference as noise, the rate region of GCF scheme $GCF_1$ contains the rate region of CF scheme, by setting $U_i = \emptyset$ in *Theorem 1* as well as the GHF scheme.

From the rate region $GCF_2$, we can see that GCF scheme strictly outperforms GHF [12] in terms of sum rate when destinations try to decode interference. In addition, the region $GCF_2$ contains the region due to CF from *Theorem 1* by setting $U_i = X_i$.

Note that in the GCF scheme, the destinations need to first decode the bin index. Therefore the relay needs to design the number of the bins according to the destination with the worst relay-destination ($R - D$) links. This is the issue for all schemes using Wyner-Ziv binning. In a concurrent work, reference [14] proposed noisy network coding, which overcomes this issue. In this scheme, the sources repeatedly transmit the same message over all blocks, and the relay simply compresses the received signal and sends the compression index to the destinations, i.e., no Wyner-Ziv binning is used. The destinations decode the source message jointly with the information received from all the blocks. The achievable rate region using noisy network coding has similar form with the achievable rate region using GCF. The improvement is in the term $R_0$ in (11) and (15) (See equations (10), (11) in [14]). For example, the second term in the sum rate expression (14) is replaced with

$$R_1 + R_2 < I(X_1 X_2 X_R; Y_i) - I(Y_R; \hat{Y}_R|X_1 X_2 X_R Y_i). \tag{22}$$

Noisy network coding in general outperforms the CF, GCF in this work and GHF in [12], at the cost of large processing delay and decoding complexity, since the same message needs to be transmitted over all blocks and joint decoding needs to be performed. Note that for the encoder, noisy network coding and GCF have similar complexity. For noisy network coding, the relay node compresses the received signal by finding a sequence from a set of i.i.d. generated sequences such that it is jointly typical with the received sequence, and then sends the index of the compression sequence to the destinations. For GCF, where Wyner-Ziv coding is utilized, the relay also needs to find the compression sequence in the same fashion. The only difference is that the relay needs to further partition the set of i.i.d. generated sequences into a number of bins and then find the bin that contains the compression sequence. The bin index is then sent to the destinations. This partition operation can be done off-line, and the relay only needs to map the compression sequence to the bin index, which does not increase the encoding complexity. As we show later in numerical examples, the performance of GCF is very close to that of the noisy network coding, despite having much less decoding complexity.

## IV. Nested Lattice Codes and GIFRC

In this section, we investigate a case where the interference link is useful in decoding the source message. We assume that the direct link is weak and the interference link is strong, and the relay uses nested lattice codes based compute-and-forward relaying. We show that this scheme can achieve higher rates than all DF and CF based relaying schemes.

Structured codes have been shown to outperform random codes in several cases [17]. Specifically, relay nodes can decode the modulo-sum of transmitted messages and forward the sum to the destinations, thus reducing the effect of the multiple access interference of the signal received at the relay. The linear structure of the codes can be exploited by both the relay and the destinations to achieve higher rates. Note that reference [18] considered multicasts in a simplified GIFRC, where there is no interference link in the channel. In [18], the relay forwards the modulo-sum of the transmitted messages to the destinations. The destinations first decode the message transmitted from the direct link, and then recover the message transmitted from the other source with the help of the modulo-sum of the messages. For our model, since the direct links are weak, we utilize the strong interference links to let the destinations decode the interference message first, and then use the signal transmitted from the relay to recover the source messages. For clarity of exposition, we consider the symmetric case, where $h_{11} = h_{22} = h_d$,

$h_{21} = h_{12} = h_c$, $h_{1R} = h_{2R} = h_s$, $h_{R1} = h_{R2} = h_R$.

*Theorem 3:* For the symmetric GIFRC, the following symmetric rate is achievable using nested lattice codes

$$R \leq \frac{1}{2}\log\left(1 + \frac{h_c^2 P}{1 + h_d^2 P}\right) \tag{23}$$

$$R \leq \frac{1}{2}\log\left(1 + \frac{h_R^2 P_R}{1 + h_d^2 P}\right) \tag{24}$$

$$R \leq \frac{1}{4}\log\left(1 + \frac{h_c^2 P + h_R^2 P_R}{1 + h_d^2 P}\right) \tag{25}$$

$$R \leq \left(\frac{1}{2}\log\left(\frac{1}{2} + h_s^2 P\right)\right)^+ \tag{26}$$

*Proof:* Since our scheme is similar to the one used in [18], here we only provide a brief summary of the encoding/decoding strategy. For preliminaries for lattice codes, see [19]. We choose a pair of nested lattice codes $\Lambda \subset \Lambda_c \subset \mathbb{R}^n$ with nesting ratio $R$, such that the coarse lattice $\Lambda$ is Rogers-good and Poltyrev-good [20], and the fine lattice $\Lambda_c$ is Poltyrev-good. We choose the coarse lattice such that $\sigma^2(\Lambda) = P$. The codewords are the fine lattice points that are within the fundamental Voronoi region of the coarse lattice. We use block Markov coding to transmit $b$ messages in $b+1$ blocks. Source $i$, $i = 1, 2$ maps its message $w_i(k)$ in block $k$ into a lattice point $t_i^n(k) \in \Lambda_c \bigcap \mathcal{V}(\Lambda)$, and transmits $X_i^n(k) = (t_i^n(k) + D_i^n(k)) \mod \Lambda$, where $D_i^n(k) \sim \text{Unif}(\mathcal{V}(\Lambda))$ is the dither. It can be shown that $X_i^n(k)$ satisfies the power constraint and is independent of $t_i^n(k)$ [19]. At the end of each block, the relay first decodes $t^n(k) = (t_1^n(k) + t_2^n(k)) \mod \Lambda$. To guarantee successful decoding, we need the constraint (26), The relay then encodes the index of this modulo-sum message into $X_R^n(k+1)$ using Gaussian signalling with power $P_R$, and transmits $X_R^n(k+1)$ to the destinations in the next block. At the destination, each decoder treats the signal from direct link, which is $X_i$ for receiver $i$, $i = 1, 2$, as noise. It then treats the signals transmitted from the relay and the interference link as a MAC. By successive decoding between the signal $X_2^n$ and $X_R^n$ and time sharing, we can show that the MAC region can be achieved, which gives us the rate constraints (23)$-$(25). ■

*Remark 2:* When direct link is weak but the interference link is strong, the information contained in the direct link is limited. Thus treating it as noise does not incur much rate loss. Instead, we can use the interference link and compute-and-forward relaying scheme to recover the message transmitted in the direct link. For the compute-and-forward relaying, we utilize the structure of the lattice codes to align the signals from different sources at the relay to *mitigate the multi-access interference* and thus removing the sum rate constraint in the MAC region. This scheme can achieve higher rates than DF based and CF based relaying schemes when the direct link is weak.

*Remark 3:* In general, when channel gains are not symmetric, the above strategy needs to be reexamined. This is because in asymmetric settings, the lattice points from two sources will not align together at the relay due to different source-relay channel gains. One possible technique to overcome this is to use channel inversion at the sources to align two lattice points together at the relay, and use the rest source power to superimpose another Gaussian signal. The performance of this scheme would suffer from the multi-access interference at the relay. A better alternative is to create a chain of nested lattice codes as described in [21], i.e., $\Lambda_1^n \subset \Lambda_2^n \subset \Lambda_C^n$ where $\sigma^2(\Lambda_1^n) = h_{1R}^2 P_1$ and $\sigma^2(\Lambda_2^n) = h_{2R}^2 P_2$, to match the different source-relay channel gains and source power. Using a chain of nested lattice codes, in this case, allows us to directly apply the strategy for the symmetric case to the asymmetric channel settings.

## V. Gaussian Interference Channel with a Potent Relay

Thus far, we have focused on achievable schemes for the GIFRC. In the sequel, we shall concentrate on deriving good outerbounds, more specifically, good sum rate upperbounds, for the GIFRC. To accomplish this task, we study the channel when the relay has infinite amount of power, and term it the GIFRC with a *potent relay*. Any capacity result for this potent relay channel serves as an outerbound for the GIFRC with finite relay power. We shall first observe the equivalence of the potent relay channel to a special case of the GIFRC, namely one with in-band reception/out-of-band noiseless transmission, which is easier to work with.

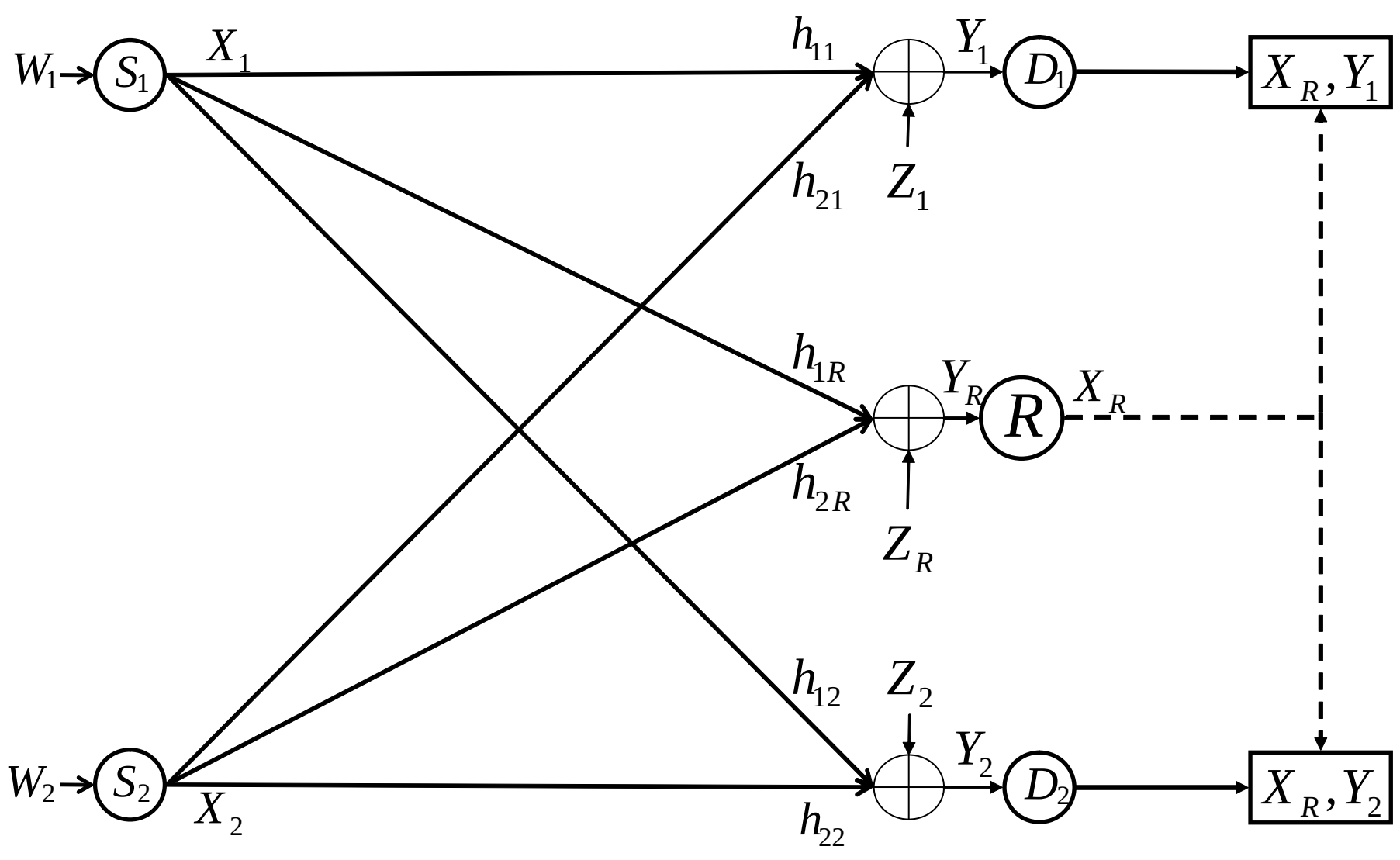

Fig. 3. GIFRC with in-band reception/out-of-band noiseless transmission.

### *A. GIFRC with in-band reception/out-of-band noiseless transmission*

Several variations of GIFRC have been studied in [3]–[7], [15]. One such variation is the channel as shown in Fig. 3. Following the notation in [15], this is the GIFRC with in-band reception/out-of-band noiseless transmission. The channel outputs are characterized by:

$$Y_1 = h_{11}X_1 + h_{21}X_2 + Z_1 \tag{27}$$

$$Y_2 = h_{12}X_1 + h_{22}X_2 + Z_2 \tag{28}$$

$$Y_R = h_{1R}X_1 + h_{2R}X_2 + Z_R \tag{29}$$

where $Z_i \sim \mathcal{N}(0,1)$ $(i = 1, 2, R)$ denotes the additive Gaussian noise at each receiver, and the channel gains are positive real numbers. Note that each destination is equipped with the signal from the relay $X_R$.

### B. *Equivalence between GIFRC with potent relay and in-band reception/out-of-band noiseless transmission*

*Proposition 1:* The capacity region ($\mathcal{C}_1$) of GIFRC with in-band reception/out-of-band noiseless transmission is asymptotically equivalent to the capacity region ($\mathcal{C}_2$) of GIFRC as the power of the relay $P_R \to \infty$.

*Proof:* $\mathcal{C}_1 \subseteq \mathcal{C}_2$: This can be shown by constructing a two stage TDMA scheme in the GIFRC, and utilizing the fact that relay has infinite power. The fraction of time allocated to the stage when relay transmit to the destination can be arbitrarily small.

$\mathcal{C}_2 \subseteq \mathcal{C}_1$: This can be shown by adding the signal $X_R$ back to the signals received at the IC with arbitrarily large gain to recover the signals with the same statistics as the one received in the IFRC with potent relay. ■

We have now established that the capacity region of GIFRC with in-band reception/out-of-band noiseless transmission is equivalent to that of the GIFRC with the potent relay. In the sequel, we will work with the former to establish the sum capacity results for the latter.

## VI. Sum Capacity of GIFRC with Potent Relay in Weak Interference

In this section, we establish the sum capacity of GIFRC with potent relay in weak interference. We proceed to work with the GIFRC with in-band reception/out-of-band noiseless transmission. In reference [16], the authors established the sum capacity of the 2-user interference channel using a “smart and useful” genie approach. Though, the application of this approach gets tedious in the general GIFRC because of the possible correlation between the codewords from the relay and those from the sources, when using a potent relay, this hardship disappears. In the following, we will upper bound our channel by SIMO interference channel with an antenna which is common to both receivers, and provide an appropriate genie information to show the optimality of Gaussian inputs. Finally, we will establish the sum capacity by achieving this upperbound by the CF based scheme. Note that for simplicity, we assume $h_{11} = h_{22} = 1$ in this section.

*Theorem 4:* For each combination of channel gains $(h_{21}, h_{12}, h_{1R}, h_{2R})$, when there exists $\rho_1, \rho_2, \rho_3, \rho_4 \in [0, 1]$ such that the following conditions hold

$$\frac{\rho_1^2}{(1+h_{21}^2P_2)^2}+\frac{h_{1R}^2\rho_3^2}{(1+h_{2R}^2P_2)^2}\geq\frac{h_{12}^2}{1-\rho_2^2}+\frac{h_{1R}^2}{1-\rho_4^2} \tag{30}$$

$$\frac{\rho_2^2}{(1+h_{12}^2P_1)^2}+\frac{h_{2R}^2\rho_4^2}{(1+h_{1R}^2P_1)^2}\geq\frac{h_{21}^2}{1-\rho_1^2}+\frac{h_{2R}^2}{1-\rho_3^2} \tag{31}$$

then the sum capacity of GIFRC with potent relay, maximum of $R_1+R_2=C_\Sigma$ is given by

$$\begin{aligned}C_\Sigma=&\frac{1}{2}\log(1+\frac{(h_{21}h_{1R}-h_{2R})^2P_1P_2+P_1+h_{1R}^2P_1}{(h_{21}^2+h_{2R}^2)P_2+1})\\&+\frac{1}{2}\log(1+\frac{(h_{12}h_{2R}-h_{2R})^2P_1P_2+h_{2R}^2P_2+P_2}{(h_{1R}^2+h_{12}^2)P_1+1})\end{aligned} \tag{32}$$

*Proof:* **Converse:** Let $S_1=h_{12}X_1+h_{12}N_1$, $S_2=h_{21}X_2+h_{21}N_2$, $S_R=h_{1R}X_1+h_{1R}N_3$, $T_R=h_{2R}X_2+h_{2R}N_4$, where $N_i\sim\mathcal{N}(0,\sigma_i^2)$, $E[N_iZ_i]=\rho_i\sigma_i$, $i=1,2$, $N_j\sim\mathcal{N}(0,\sigma_j^2)$, $E[N_jZ_R]=\rho_j\sigma_j$, $j=3,4$. These are the genie information we shall utilize. We have that

$$\begin{aligned}&n(R_1+R_2)\\&\leq I(W_1;\hat{W}_1)+I(W_2;\hat{W}_2) &(33)\\&\leq I(X_1^n;Y_1^n,Y_R^n)+I(X_1^n;X_R^n|Y_1^n,Y_R^n)+I(X_2^n;Y_2^n,Y_R^n)+I(X_2^n;X_R^n|Y_2^n,Y_R^n) &(34)\\&=I(X_1^n;Y_1^n,Y_R^n)+I(X_2^n;Y_2^n,Y_R^n) &(35)\\&\leq I(X_1^n;Y_1^n,Y_R^n,S_1^n,S_R^n)+I(X_2^n;Y_2^n,Y_R^n,S_2^n,T_R^n) &(36)\\&=h(S_1^n,S_R^n)-h(S_1^n,S_R^n|X_1^n)+h(Y_1^n,Y_R^n|S_1^n,S_R^n)-h(Y_1^n,Y_R^n|S_1^n,S_R^n,X_1^n)\\&\quad+h(S_2^n,T_R^n)-h(S_2^n,T_R^n|X_2^n)+h(Y_2^n,Y_R^n|S_2^n,T_R^n)-h(Y_2^n,Y_R^n|S_2^n,T_R^n,X_2^n) &(37)\end{aligned}$$

here, (35) is due to the Markov chain $X_1^i,X_2^i\to Y_R^i\to X_{R,i+1}$. In inequality (36), we give the genie information to both the receivers. Then (37) can be written as

$$\begin{aligned}&h(h_{12}X_1^n+h_{12}N_1^n,h_{1R}X_1^n+h_{1R}N_3^n)-h(h_{21}X_2^n+Z_1^n,h_{2R}X_2^n+Z_R^n|N_1^n,N_3^n)\\&+h(X_1^n+h_{21}X_2^n+Z_1^n,h_{1R}X_1^n+h_{2R}X_2^n+Z_R^n|h_{12}X_1^n+h_{12}N_1^n,h_{1R}X_1^n+h_{1R}N_3^n)\end{aligned}$$

$$+ h(h_{21}X_2^n + h_{21}N_2^n, h_{2R}X_2^n + h_{2R}N_4^n) - h(h_{12}X_1^n + Z_2^n, h_{1R}X_1^n + Z_R^n|N_2^n, N_4^n)$$

$$+ h(h_{12}X_1^n + X_2^n + Z_2^n, h_{1R}X_1^n + h_{2R}X_2^n + Z_R^n|h_{21}X_2^n + h_{21}N_2^n, h_{2R}X_2^n + h_{2R}N_4^n)$$

$$- h(h_{12}N_1^n, h_{1R}N_3^n) - h(h_{21}N_1^n, h_{2R}N_4^n) \tag{38}$$

To guarantee that i.i.d. Gaussian inputs maximize (38), we need the following terms to be maximized by Gaussian inputs, which is stated in *Lemma 1*.

$$h(h_{12}X_1^n + h_{12}N_1^n, h_{1R}X_1^n + h_{1R}N_3^n) - h(h_{12}X_1^n + Z_2^n, h_{1R}X_1^n + Z_R^n|N_2^n, N_4^n) \tag{39}$$

$$h(h_{21}X_2^n + h_{21}N_2^n, h_{2R}X_2^n + h_{2R}N_4^n) - h(h_{12}X_2^n + Z_1^n, h_{2R}X_2^n + Z_R^n|N_1^n, N_3^n) \tag{40}$$

*Lemma 1:* When there exist $\sigma_1^2, \sigma_2^2, \sigma_3^2, \sigma_4^2 \geq 0$ and $\rho_1, \rho_2, \rho_3, \rho_4 \in [0, 1]$ such that the following condition holds

$$\frac{1}{\sigma_1^2} + \frac{1}{\sigma_3^2} \geq \frac{h_{12}^2}{1-\rho_2^2} + \frac{h_{1R}^2}{1-\rho_4^2} \tag{41}$$

$$\frac{1}{\sigma_2^2} + \frac{1}{\sigma_4^2} \geq \frac{h_{21}^2}{1-\rho_1^2} + \frac{h_{2R}^2}{1-\rho_3^2} \tag{42}$$

Then i.i.d. Gaussian inputs with variance $P_1$ and $P_2$ maximize (39) and (40).

For the proof of *Lemma 1*, see Appendix C.

It then follows that the expression (36) is equivalent to

$$nI(X_{1G}; Y_1, Y_R, S_1, S_R) + nI(X_{2G}; Y_2, Y_R, S_2, T_R) \tag{43}$$

where $X_{iG} \sim \mathcal{N}(0, P_i),\ i = 1, 2$. Here, $X_{iG}$ represents the i.i.d. Gaussian inputs. Next, we show how to make the genie that supplies $S_1, S_2, S_R, T_R$ “smart”.

*Lemma 2:* Under the conditions

$$\rho_1\sigma_1 = 1 + h_{21}^2 P_2 \qquad \rho_2\sigma_2 = 1 + h_{12}^2 P_1$$

$$c\rho_3\sigma_3 = 1 + h_{2R}^2 P_2 \qquad d\rho_4\sigma_4 = 1 + h_{1R}^2 P_1 \tag{44}$$

the genie is also *smart* in the sense that

$$I(X_{1G};Y_1,Y_R,S_1,S_R)+I(X_{2G};Y_2,Y_R,S_2,T_R)=I(X_{1G};Y_1,Y_R)+I(X_{2G};Y_2,Y_R) \tag{45}$$

For the proof of *Lemma 2*, see Appendix D.

Then, using *Lemma 2* and *Lemma 1*, the sum rate can be bounded by

$$R_1+R_2\leq I(X_{1G};Y_1,Y_R)+I(X_{2G};Y_2,Y_R) \tag{46}$$

which gives us the expression (32).

**Achievability:** When we evaluate the rate region $GCF_1$ in section III by Gaussian inputs with $X_1\sim\mathcal{N}(0,P_1)$,$X_2\sim\mathcal{N}(0,P_2)$,$X_R\sim\mathcal{N}(0,P_R)$,$\hat{Y}_R=Y_R+\hat{Z}$, $\hat{Z}\sim\mathcal{N}(0,\sigma_R^2)$, we can show that the sum rate expression reduces to (32) when $P_R\to\infty$. ■

*Remark 4:* When the conditions in *Lemma 1* do not hold, it is also possible to bound the sum rate using this set of genie information. However, the bound is loose. The reason is that we are maximizing the terms in (38) separately. In this case, the power that maximizes (39) and (40) is 0, while the power that maximizes other terms in (38) is $P_1$ for $X_1$, and $P_2$ for $X_2$. By contrast, when conditions in *Lemma 1* hold, the power that maximizes the terms in (38) is the same.

*Proposition 2:* For the symmetric case, when $\rho_1=\rho_2=\rho$, $\rho_3=\rho_4=\rho_R$, the conditions (30) and (31) are equivalent to

$$h_s^2\leq\frac{1-2h_c(1+h_c^2P)}{1+h_c^2P} \tag{47}$$

*Proof:* For the symmetric case, (30) and (31) are the same. The expression (30) can be written as

$$\frac{\rho^2}{(1+h_c^2P)^2}-\frac{h_c^2}{1-\rho^2}+\frac{h_s^2\rho_R^2}{(1+h_s^2P)^2}-\frac{h_s^2}{1-\rho_R^2}\geq 0 \tag{48}$$

We can find $\rho,\rho_R\in[0,1]$ such that the above condition is satisfied if and only if the maximum of the left hand side of the inequality is greater than or equal to 0. Observe that

$$\frac{\rho^2}{(1+h_c^2P)^2}-\frac{h_c^2}{1-\rho^2} \tag{49}$$

is a concave function of $\rho^2$ and

$$\frac{h_s^2\rho_R^2}{(1+h_s^2P)^2} - \frac{h_s^2}{1-\rho_R^2} \tag{50}$$

is also concave in $\rho_R^2$. Thus, it is easy to see that $\rho^{2*} = 1 - h_c(1+h_c^2P)$ and $\rho_R^{2*} = 0$ are the maximizer. Then (9) reduces to

$$h_s^2 \leq \frac{1-2h_c(1+h_c^2P)}{1+h_c^2P} \tag{51}$$

■

*Remark 5:* The examination of the symmetric case gives us insight of what range of channel gains conditions (30) and (31) imply. First, the interference links should be weak. This can be seen from $1-2h_c(1+h_c^2P) \geq 0$. Also, the $S-R$ links should not be strong, i.e. $h_s^2 \leq 1$.

## VII. Sum Capacity of GIFRC with Potent Relay in Strong Interference

In this section, we shall establish the strong interference condition following the method in [8], [22], [23], under which the channel capacity of GIFRC with potent relay can be found. Similar to the weak interference case in section VI, we proceed to work with the GIFRC with in-band reception/out-of-band noiseless transmission.

*Theorem 5:* When $h_{12} \geq h_{11}$ and $h_{21} \geq h_{22}$, the capacity region of GIFRC with potent relay is

$$\begin{aligned}
&R_1 \leq \frac{1}{2}\log(1+h_{11}^2P_1+h_{1R}^2P_1)\\
&R_2 \leq \frac{1}{2}\log(1+h_{22}^2P_2+h_{2R}^2P_2)\\
&R_1+R_2 \leq \frac{1}{2}\log(1+\min\{(h_{21}h_{1R}-h_{11}h_{2R})^2P_1P_2+(h_{1R}^2+h_{11}^2)P_1+(h_{2R}^2+h_{21}^2)P_2,\\
&\qquad (h_{12}h_{2R}-h_{1R}h_{22})^2P_1P_2+(h_{1R}^2+h_{12}^2)P_1+(h_{2R}^2+h_{22}^2)P_2\}) 
\end{aligned} \tag{52}$$

*Proof:* From *Proposition 1*, we focus on the channel with in-band reception/out-of-band noiseless transmission. Based on the techniques bounding the strong interference channel in [8], with $h_{12} \geq h_{11}$ and $h_{21} \geq h_{22}$, we first assume decoder 1 and decoder 2 can decode their own messages. For decoder 1, with $Y_1^n = h_{11}X_1^n + h_{21}X_2^n + Z_1^n$ and $X_R^n$, by constructing $h_{22}\frac{Y_1^n - h_{11}X_1^n}{h_{21}} + h_{12}X_1^n + N_1^n$, where $N_1 \sim \mathcal{N}(0, \sqrt{1 - \frac{h_{22}^2}{h_{21}^2}})$, it can also decode $w_2$. Similar result can be obtained for decoder 2 in the same way. It follows that any code for the GIFRC channel with potent relay is also a code for the compound SIMO MAC channel with an antenna common to both receivers.

From (35), we can outer bound the capacity using techniques for the MAC channel. For the achievability, When we evaluate the rate region $GCF_2$ in section III by Gaussian inputs with $X_1 \sim \mathcal{N}(0, P_1)$,$X_2 \sim \mathcal{N}(0, P_2)$,$X_R \sim \mathcal{N}(0, P_R)$,$\hat{Y}_R = Y_R + \hat{Z}$, $\hat{Z} \sim \mathcal{N}(0, \sigma_R^2)$, we can show that the rate region reduces to the rate region in *Theorem 5* when $P_R \to \infty$. ■

## VIII. Degrees of Freedom of the GIFRC

In this section, we characterize the degrees of freedom ($DoF$) of the GIFRC. The $DoF$ is defined as

$$DoF = \lim_{P\to\infty} \frac{C_{sum}}{\frac{1}{2}\log P} \tag{53}$$

where $C_{sum}$ is the sum capacity of the channel, $P = P_1 = P_2$ is the source power, and we assume the noise power is unity.

We observe that the $DoF$ depends on how fast the relay power grows in relation to the power of the sources.

*Proposition 3:* The $DoF$ of the GIFRC is $1$ when $P_R = \mathcal{O}(P)$, while the $DoF$ of the GIFRC is $2$ when $P_R = \mathcal{O}(P^k), k \geq 2$, as $P \to \infty$.

*Proof:* For the case when $P_R = \mathcal{O}(P)$, we can combine the relay with one source, and the channel becomes the MIMO interference channel with cooperation between the sources, where one transmitter has two antennas and the other transmitter and receivers have one antenna. From *Corollary 11* in [24], this approach indicates that the upperbound for the $DoF$ for the GIFRC is $1$. Random coding argument, e.g., the one in [11], achieves this $DoF$.

For the case when $P_R = \mathcal{O}(P^k), k \geq 2$, we first consider the case when the relay is potent. For the SIMO interference channel where transmitters have one antenna and receivers have two antennas, the $DoF$ is $2$ [25]. This provides an upperbound for the $DoF$ for the GIFRC with potent relay. By evaluating the rate expressions in $CF_2$ with $X_1 \sim \mathcal{N}(0, P)$,$X_2 \sim \mathcal{N}(0, P)$,$X_R \sim \mathcal{N}(0, P_R)$,$\hat{Y}_R = Y_R + \hat{Z}$, $\hat{Z} \sim \mathcal{N}(0, \sigma_R^2)$, we can see that CF scheme achieves this upperbound as $P \to \infty$, under the condition that the relay is *potent* in the first place. Thus, the $DoF$ for the GIFRC with potent relay is 2. By further evaluation of the rates achieved by the CF scheme, i.e., (5) – (8), we can see that, in fact, this $DoF$ can be achieved by the general GIFRC when the power of the relay satisfies $P_R = \mathcal{O}(P^k), k \geq 2$, as $P \to \infty$. ■

*Remark 6:* Reference [26] showed that for the interference channel, using a MIMO relay with power proportional to $\mathcal{O}(P^2)$, the $DoF$ of 2 is achievable. Our result indicates that the relay does not need to have multiple antennas to achieve the $DoF$ of this channel.

## IX. Cut-set Bound

In this section, we provide the cut-set bound for the GIFRC and compare it with our potent relay outerbound.

*Proposition 4:* The following rate region is an outerbound for the IFRC

$$\mathcal{R}_{cutset} = \bigcup_{\substack{p(x_1)p(x_2) \\ p(x_R|x_1x_2)}} \mathcal{R}$$

where $\mathcal{R}$ is the set of rate pairs $(R_1, R_2)$ satisfying

$$R_1 \leq \min\{I(X_1X_R; Y_1|X_2), I(X_1; Y_1Y_R|X_2X_R)\} \tag{54}$$

$$R_2 \leq \min\{I(X_2X_R; Y_2|X_1), I(X_2; Y_2Y_R|X_1X_R)\} \tag{55}$$

$$R_1 + R_2 \leq \min\{I(X_1X_2; Y_1Y_2Y_R|X_R), I(X_1X_2X_R; Y_1Y_2)\} \tag{56}$$

for one specific distribution $p(x_1)p(x_2)p(x_R|x_1x_2)$.

For the Gaussian channel when $Y_1, Y_2, Y_R$ satisfy the equations (2) to (4), it is obvious that Gaussian inputs satisfying the power constraint maximize all the three terms in the cut-set bound. Now, we focus on the sum rate bound (56). This bound still needs to be maximized over the correlation coefficients $\rho_{R1}, \rho_{R2}$ between $X_R$ and $X_1, X_2$. We claim that our potent relay outerbound is at least tighter than the first term in the sum rate bound (56). To see this, first we notice that for the Gaussian inputs $X_{1G}, X_{2G}, X_{RG}$,

$$I(X_{1G}X_{2G}; Y_{1G}Y_{2G}Y_{RG}|X_{RG}) = I(X_{1G}X_{2G}; Y'_{1G}Y'_{2G}Y'_{RG}|X_{RG}) \tag{57}$$

where $Y'_{1G}, Y'_{2G}, Y'_{RG}$ satisfy (27) to (29).

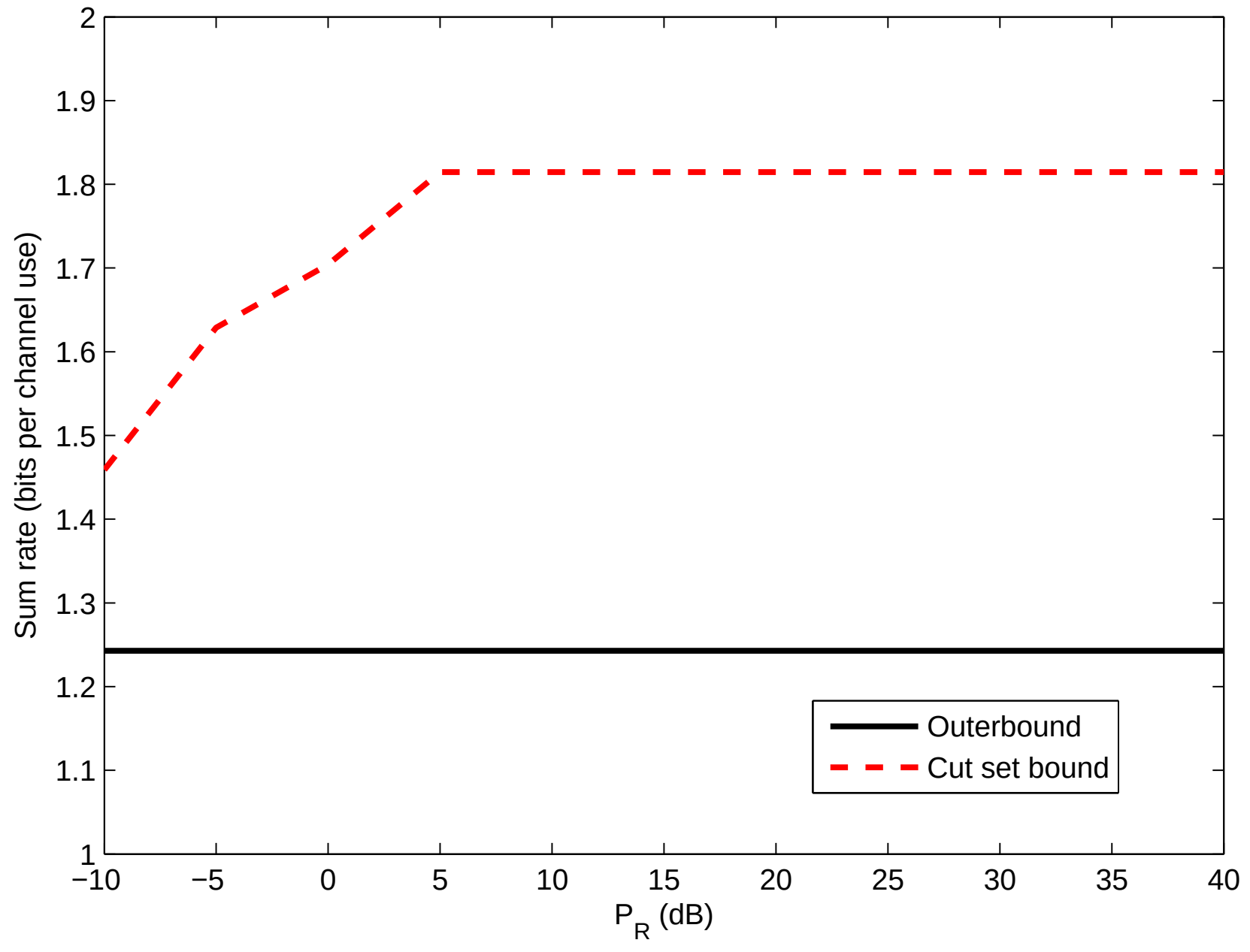

Fig. 4. Comparison of the cut-set bound and our bound under weak interference when $P = 1\text{dB}, h_d = 1, h_c = \sqrt{0.1}$, $h_s = \sqrt{0.8}, h_R = 1$.

For the case of strong interference, our potent relay sum rate upperbound is

$$\min\{I(X_{1G}X_{2G}; Y'_{1G}Y'_{RG}), I(X_{1G}X_{2G}; Y'_{2G}Y'_{RG})\} \tag{58}$$

We have

$$\max_{\rho_{R1},\rho_{R2}} I(X_{1G}X_{2G}; Y'_{1G}Y'_{2G}Y'_{RG}|X_{RG}) \tag{59}$$

$$= I(X_{1G}X_{2G}; Y'_{1G}Y'_{2G}Y'_{RG}) \tag{60}$$

$$\geq I(X_{1G}X_{2G}; Y'_{1G}Y'_{RG}) \tag{61}$$

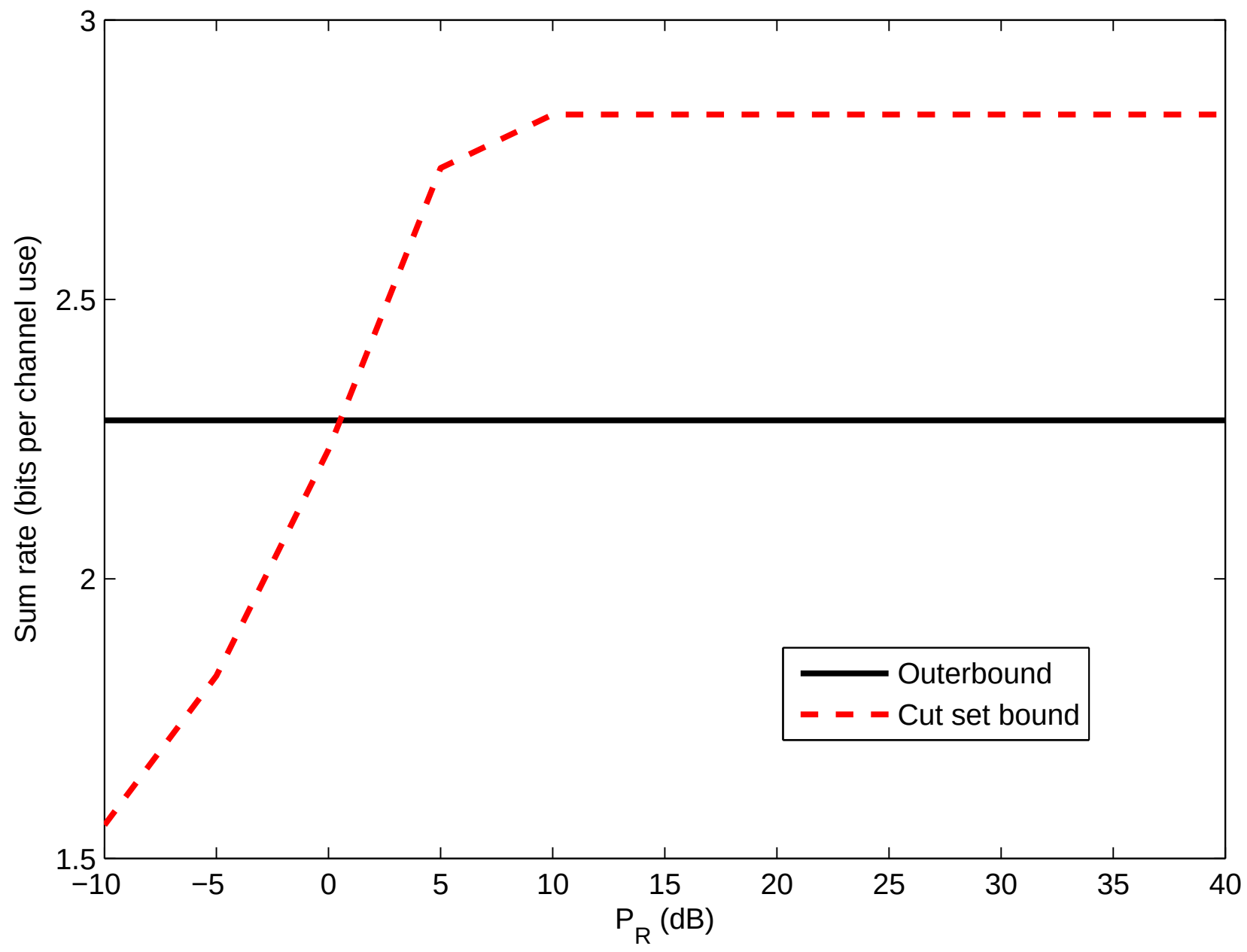

Fig. 5. Comparison of the cut-set bound and our bound under strong interference when $P = 1\text{dB}, h_d = 1, h_c = 2,\ h_s = 2, h_R = 1$.

Similarly, the sum rate cut-set bound is greater than or equal to $I(X_{1G}X_{2G}; Y'_{2G}Y'_{RG})$.

For the case of weak interference, potent relay sum rate upperbound is $I(X_{1G}; Y'_{1G}Y'_{RG}) + I(X_{2G}; Y'_{2G}Y'_{RG})$. We have

$$\max_{\rho_{R1},\rho_{R2}} I(X_{1G}X_{2G}; Y'_{1G}Y'_{2G}Y'_{RG}|X_{RG}) \tag{62}$$

$$= I(X_{1G}X_{2G}; Y'_{1G}Y'_{2G}Y'_{RG}) \tag{63}$$

$$\geq I(X_{1G}; Y'_{1G}Y'_{2G}Y'_{RG}) + I(X_{2G}; Y'_{1G}Y'_{2G}Y'_{RG}) \tag{64}$$

$$\geq I(X_{1G}; Y'_{1G}Y'_{RG}) + I(X_{2G}; Y'_{2G}Y'_{RG}) \tag{65}$$

Figure 4 and 5 show the comparison between the potent relay outerbounds and the cut set bounds as a function of the power of the relay for weak and strong interference. We can see that the potent relay outerbound is tighter than the cut set bound even when the power of the relay is moderate.

*Remark 7:* In a recent paper [27], the authors proposed another outerbound for GIFRC, which is at least tighter than the second term in the sum rate cut-set bound. The bound in [27] thus can be thought of one that complements our bound. Our bound is tighter when the relay has moderate and large power, whereas the bound in [27] is tighter when the relay has small power.

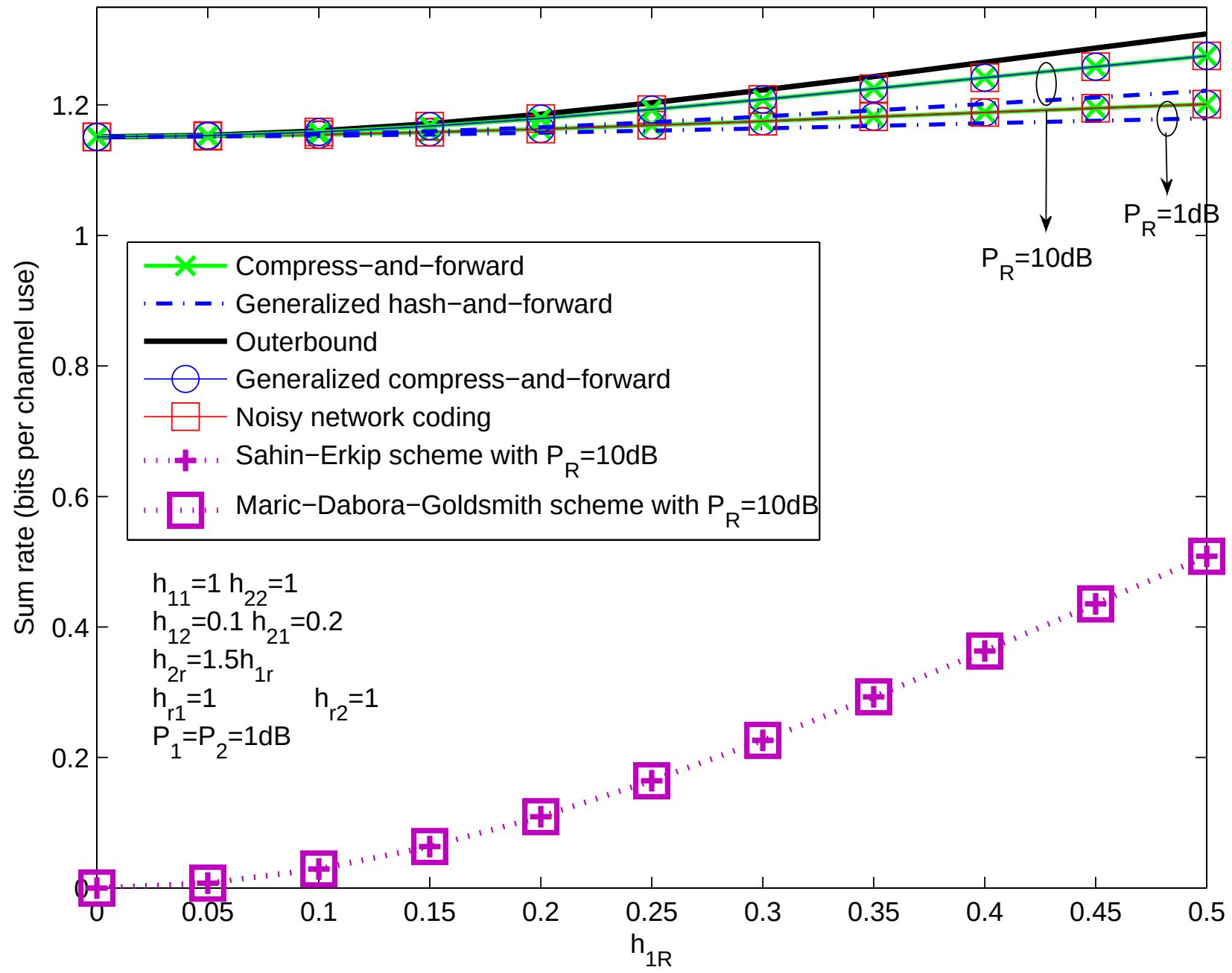

Fig. 6. Comparison of the potent relay outerbounds and achievable sum rates under weak interference.

## X. Numerical Results

In this section, by numerical results, we compare the achievable sum rates with the potent relay upperbounds we derived in previous sections.

Figure 6 compares the potent relay outerbound (32) and the achievable sum rates due to compress-and-forward (CF), generalized hash-and-forward (GHF) [12], generalized compress-and-forward (GCF) and noisy network coding (NNC) [14] for weak interference. We also plot the performance of two DF based scheme from [3], [11]. We term the achievable scheme from [3] the Sahin-Erkip scheme, and the one from [11] the Maric-Dabora-Goldsmith scheme. The channel parameters are shown on the figure. Specifically, the achievable sum rates are obtained from $GCF_1$ in section III and from *Theorem 1* by treating interference as noise. Note that the constraints (30) and (31) require both the interference links and source-relay links to be weak, and thus the DF based relaying scheme does not perform well. CF, GCF and NNC have similar performance and outperform GHF. We can see that when the power of the relay is 10dB, the achievable sum rates are very close to the potent relay outerbound, with a gap less than 0.05 bits per channel use.

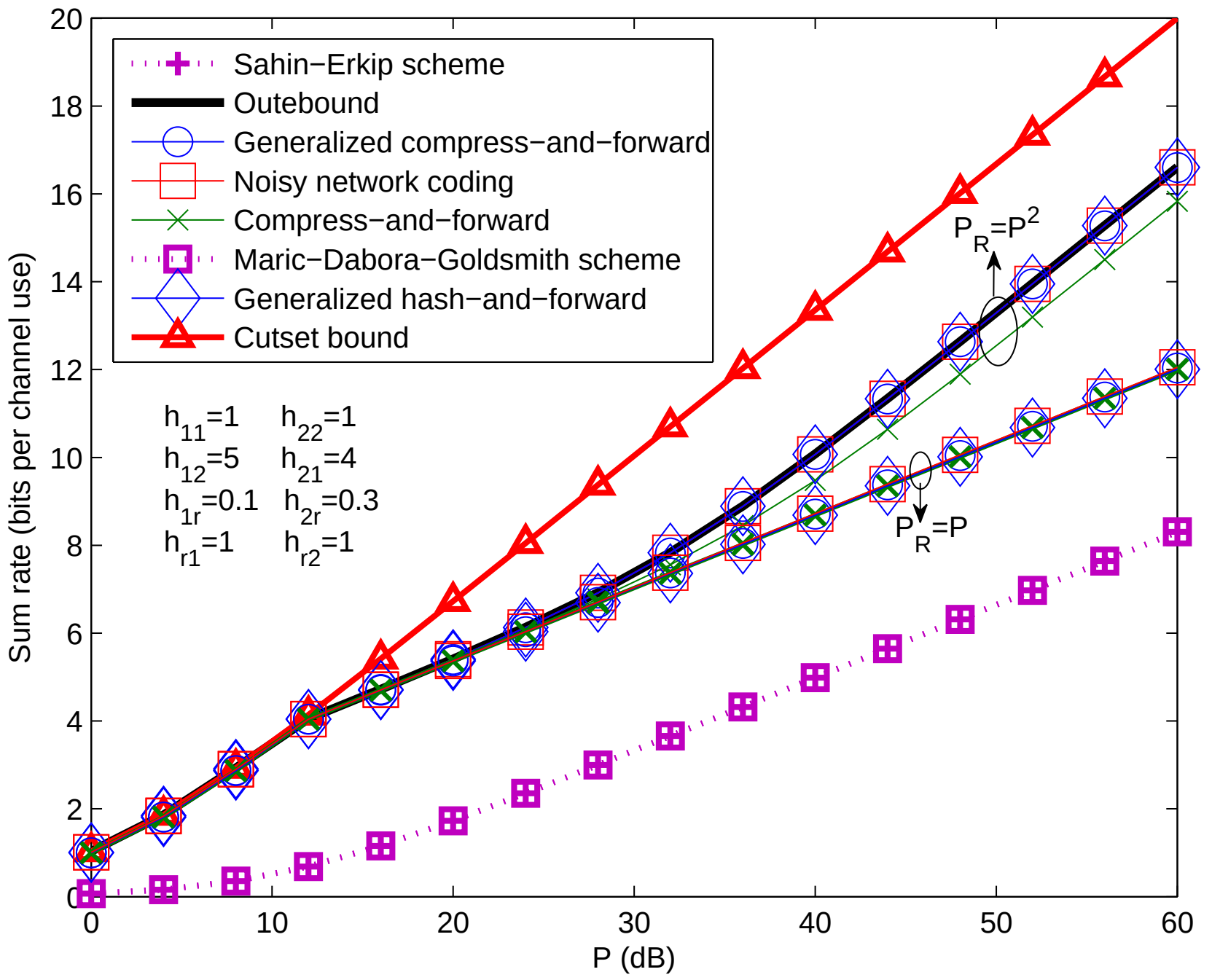

Fig. 7. Comparison of the potent relay outerbounds and achievable sum rates under strong interference with $P_1 = P_2 = P$.

Figure 7 shows the potent relay outerbound from *Theorem 5* and the achievable sum rates for strong interference. The channel parameters are shown on the figure. We can see that the potent relay outerbound is tighter than the cutset bound. The rates for CF and GCF are based on *Theorem 1* by decoding interference and $GCF_2$, respectively. When the power of the relay is of the same order as that of the sources, i.e., $P_R = \mathcal{O}(P)$, the outerbound and the achievable sum rates coincide numerically for low to moderate power values. When the power of the relay is of the order of $\mathcal{O}(P^2)$ of the sources, or $P_R = 2P$ in dB, the achievable sum rates and the potent relay outerbound coincide numerically for all power values. This shows that the potent relay

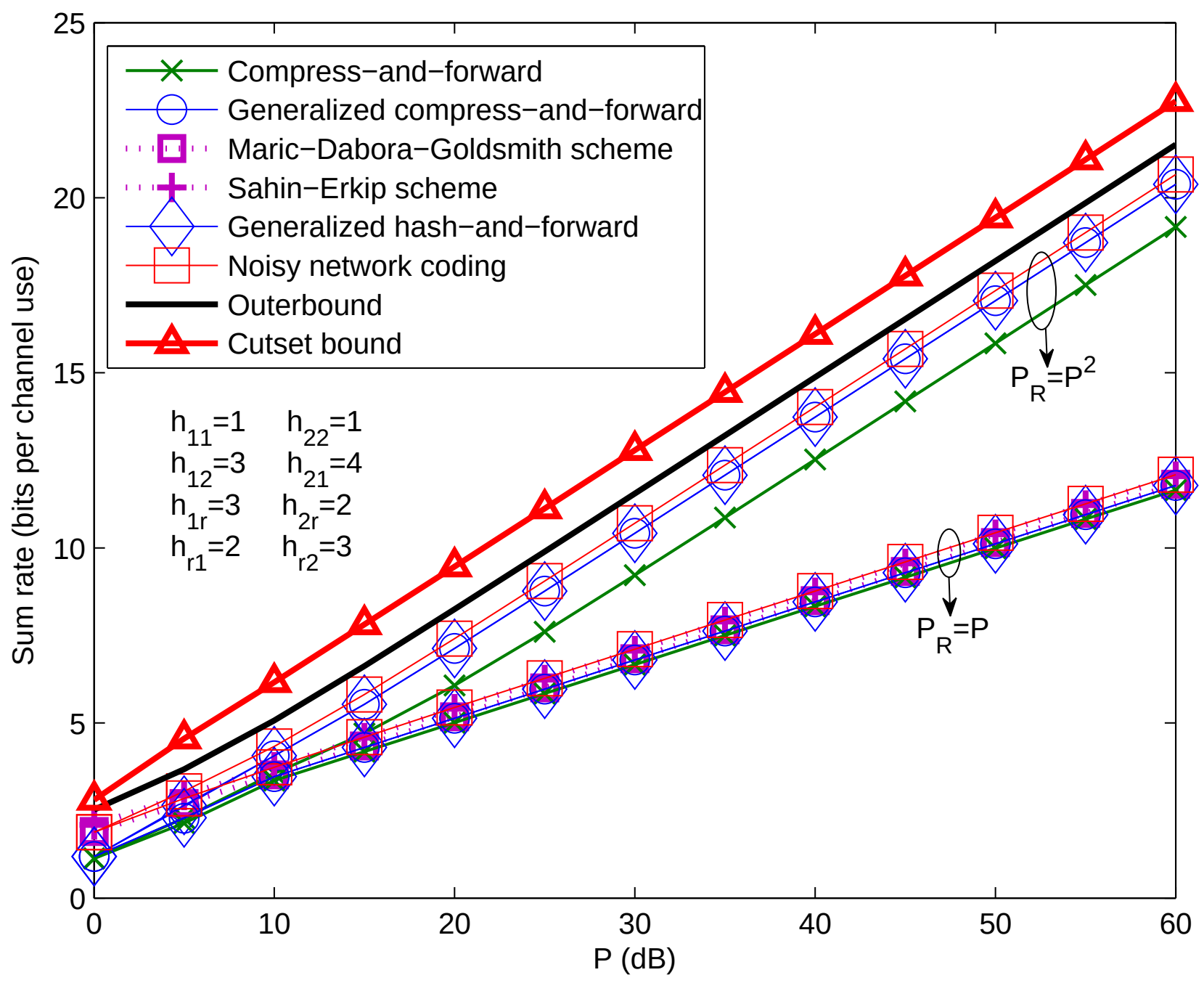

Fig. 8. Comparison of the potent relay outerbounds and achievable sum rates under strong interference and source-relay links and asymmetry in the relay-destination links with $P_1 = P_2 = P$.

outerbound is tight when the relay has large power compared with the power of the received signal, but the relay power does not need to be infinite. In addition, we can see that GHF, GCF and NNC have similar performance, while both improving the rates achieved by CF relaying. Figure 8 shows the comparison of the outerbounds and achievable rates for strong source-relay links and asymmetric relay-destination links. The potent relay outerbound is once again tighter than the cutset bound for this range of channel parameters and is close to the achievable rates with a small constant gap when $P_R = \mathcal{O}(P^2)$. When $P_R = \mathcal{O}(P)$, DF type of relaying strategies perform better than CF type of relaying strategies. However, when $P_R = \mathcal{O}(P^2)$, the performance of DF type of relaying strategies is limited by the source power, and has the same performance as the case when $P_R = \mathcal{O}(P)$. Therefore the CF type of strategies have better performance for the case $P_R = \mathcal{O}(P^2)$. In addition, we can see that when the relay-destination links are asymmetric, the improvement of NNC upon GCF scheme is very limited, while both improving

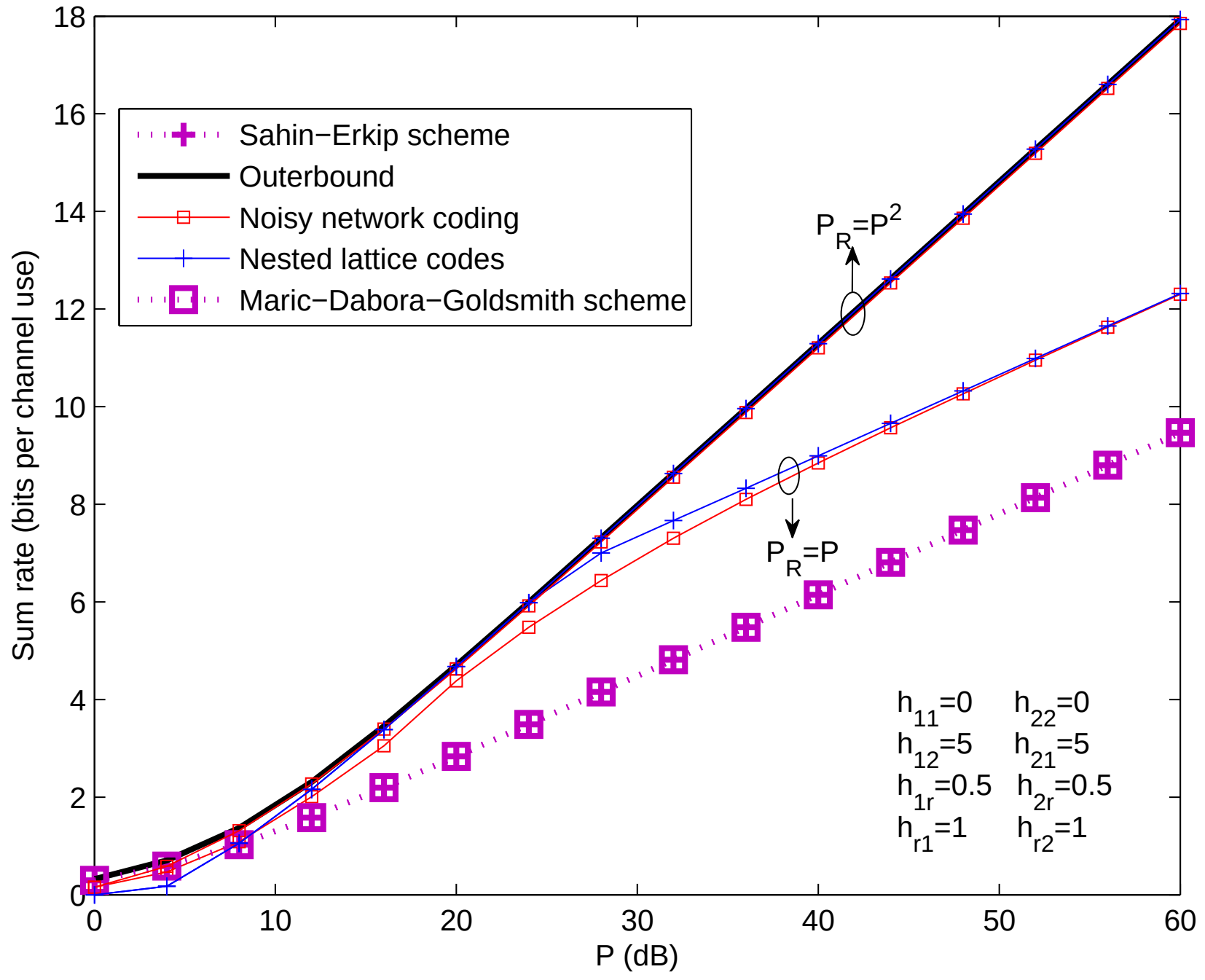

Fig. 9. Comparison of the potent relay outerbounds and achievable sum rates under strong interference and weak direct link with $P_1 = P_2 = P$.

the rates achieved by the CF scheme.

Figure 9 shows the potent relay outerbound from *Theorem* 5 and the achievable sum rates for strong interference and weak direct link. The channel parameters are shown on the figure. Since the direct link is weak, CF, GHF and GCF do not perform as well as NNC since they rely on Wyner-Ziv binning, and the side information $Y_1^n$ ($Y_2^n$) contains little information about the messages transmitted through the direct links. We thus only compare NNC and the lattice code based compute-and-forward relaying in section IV. We can see that when $P_R = \mathcal{O}(P)$, lattice code based compute-and-forward relaying outperforms NNC. Both schemes coincide with the potent relay outerbound when $P_R = \mathcal{O}(P^2)$, or $P_R = 2P$ in dB.

Figure 7, 8 and 9 also illustrate the $DoF$ of the IFRC. When the relay has large power compared with the sources, the $DoF$ of the channel increases from 1 to 2, which is illustrated by the slope of the curves. CF based relaying (CF, GCF and NNC) and lattice code based compute-and-forward relaying achieves the $DoF$, while DF based relaying strategies can only achieve $DoF$ of 1 even when the relay has large power, due to the constraints of the decoding capability at the relay.

## XI. Conclusion

In this paper, we have studied the Gaussian interference channel with an intermediate relay. We have proposed a compress-and-forward (CF) relaying scheme, which requires the destinations to uniquely decode the compression index. We have also proposed a GCF relaying scheme, where the destinations do not need to decode and compression indices. We have shown that the GCF scheme outperforms the CF scheme, and the recently proposed generalized hash-and-forward strategy. We have also designed a nested lattice code based compute-and-forward relaying strategy, which outperforms all the existing strategies, including noisy network coding, when direct link is weak and interference link is strong. In fact, this scheme shows that the interference link can be useful in decoding the source messages with the help of a relay. We have also devised a new and useful sum capacity upper bound for this channel. We have accomplished this by examining the channel when the relay has very large power, i.e. the GIFRC with a potent relay. We have found the sum capacity of this channel under weak and strong interference conditions. Both results serve, in turn, as sum rate upperbounds for GIFRC with finite relay power constraint. We observe that the bound is useful in the sense that it is close to the known achievable schemes under channel conditions in most scenarios of interest. We conclude by noting that although the capacity region of the IFRC in general remains elusive, attempts towards useful upperbounds and achievable strategies provide us with helpful insights for designing interference networks.

## Appendix A
## Proof of Theorem 1

*Proof:* We use block Markov encoding [10].

*Codebook for the sources:* Choose a joint distribution

$$p(u_1)p(u_2)p(x_1|u_1)p(x_2|u_2)p(x_R)p(y_1y_2y_R|x_1x_2x_R)p(\hat{y}_R|y_Rx_R)$$

Split message $w_i$ into $w_{i0}, w_{ii}$ where $w_{i0} \in \{1, \ldots, 2^{nR_{i0}}\}$, $w_{ii} \in \{1, \ldots, 2^{nR_{ii}}\}$, and $R_{i0} + R_{ii} = R_i$, $i = 1, 2$. For each $w_{i0}$, generate the codeword $u_i^n(w_{i0})$ according to $p(u_i^n(w_{i0})) = \prod_{k=1}^n p(u_{i,k})$. For each $u_i^n(w_{i0})$, generate $2^{nR_{ii}}$ codewords $x_i^n(w_{ii}|w_{i0})$ for each message $w_{ii}$ according to $p(x_i^n(w_{ii}|w_{i0})) = \prod_{k=1}^n p(x_{i,k}|u_{i,k}(w_{i0}))$.

*Codebook for the relay:* Choose $2^{nR_0}$ codewords $x_R^n(w_0)$ for $w_0 \in \{1, \ldots, 2^{nR_0}\}$ according to $p(x_R^n(w_0)) = \prod_{k=1}^n p(x_{R,k}(w_0))$. Then, for each $x_R^n(w_0)$, choose $2^{n\hat{R}}$ codewords $\hat{y}_R^n(z|w_0)$for each $z \in \{1, \ldots, 2^{n\hat{R}}\}$ according to $p(\hat{y}_R^n(z|w_0)) = \prod_{k=1}^n p(\hat{y}_{R,k}|x_{R,k}(w_0))$. Randomly partition the set $\{1, \ldots, 2^{n\hat{R}}\}$ into $2^{nR_0}$ cells $\{S(w_0)\}$.

*Encoding:* In block $k$, let $w_{10}(k), w_{11}(k), w_{20}(k), w_{22}(k)$ and $w_0(k), z(k)$ be the messages to be sent from the sources and the relay, respectively. For the sources, choose the corresponding codewords $u_1^n(w_{10}(k)), x_1^n(w_{11}(k)|w_{10}(k))$, $u_2^n(w_{20}(k)), x_2^n(w_{22}(k)|w_{10}(k))$ to be sent in this block. For the relay, assume that $(\hat{y}_R^n(z(k-1)|w_0(k-1)), y_R^n(k-1), x_R^n(w_0(k-1))) \in T_\epsilon$, where $T_\epsilon$ stands for the jointly $\epsilon$-typical set, and $z(k-1) \in S(w_0(k))$, then $x_R^n(w_0(k))$ is transmitted in block $k$.

*Decoding:* At the end of block $k$, two receivers decode $w_0(k)$ to obtain an estimate $\hat{w}_0(k)$ independently. To successfully decode $w_0(k)$, we need $R_0 \leq \min\{I(X_R; Y_1), I(X_R; Y_2)\}$. Then both receivers try to find $\hat{z}(k-1)$ such that $(\hat{y}_R^n(\hat{z}(k-1)|\hat{w}_0(k-1)), y_i^n(k-1), x_R^n(\hat{w}_0(k-1))) \in T_\epsilon$ and $\hat{z}(k-1) \in S(w_0(k))$. To successfully decode this, we need $\hat{R} \leq \min\{I(\hat{Y}_R; Y_1|X_R) + I(X_R; Y_1), I(\hat{Y}_R; Y_2|X_R) + I(X_R; Y_2)\}$.

After the relay correctly decodes the quantized version of the signal it received in block $k-1$, decoder 1 tries to find $(w_{10}(k-1), w_{11}(k-1), w_{20}(k-1))$ such that

$$\Big(u_1^n(w_{10}(k-1)), x_1^n(w_{11}(k-1)|w_{10}(k-1)), u_2^n(w_{20}(k-1)),$$
$$\hat{y}_R^n(\hat{z}(k-1)|\hat{w}_0(k-1)), y_1^n(k-1), x_R^n(\hat{w}_0(k-1))\Big) \in T_\epsilon$$

Decoder 2 uses the same method to decode $(w_{20}(k-1), w_{22}(k-1), w_{10}(k-1))$. The rate region can be obtained to guarantee vanishing error probability. ■

# APPENDIX B
## PROOF OF THEOREM 2

*Proof:* We only show the proof of the rate region $GCF_2$, i.e., the destinations try to decode the interference. $GCF_1$ follows from similar steps.

*Codebook generation:* Fix a distribution $p(x_1)p(x_2)p(x_R)p(\hat{y}_R|x_R y_R)$. For each message $w_i \in \{1, \cdots, 2^{nR_i}\}$, generate codeword $x_i^n(w_i)$ according to $\prod_{j=1}^n p_{X_i}(x_{ij})$ $(i = 1, 2)$. For each $k \in \{1, \cdots, 2^{nR_0}\}$ generate $x_R^n(k)$ according to $\prod_{j=1}^n p_{X_R}(x_{Rj})$. For each$x_R^n(k)$, generate $\hat{y}_R^n(l|k)$ for each $l \in \{1, \cdots, 2^{n\hat{R}}\}$ according to $\prod_{j=1}^n p_{\hat{y}_R|X_R}(\hat{y}_{Rj}|x_{Rj})$. Partition the set $\{1, \cdots, 2^{n\hat{R}}\}$ into $2^{nR_0}$ bins.

*Encoding:* We use block Markov encoding for $B$ blocks. For block $b$, source 1 encodes $w_{1b}$ into $x_1^n(w_{1b})$ and sources 2 encodes $w_{2b}$ into $x_2^n(w_{2b})$. The relay receives $y_{R,b}^n$, and it looks for $l_b \in \{1, \cdots, 2^{n\hat{R}}\}$ such that $(y_{R,b}^n, \hat{y}_R^n(l_b|k_{b-1}), x_R^n(k_{b-1})) \in T_\epsilon^n$. If no such $l_b$ exists, the relay declares an error. If there is more than one such $l_b$, the relay chooses the smallest one. Note that $l_b \in \mathcal{B}(k_b)$. The relay sends $x_R^n(k_b)$ in block $b+1$. Note that we fix $w_{1B} = 1$, $w_{2B} = 1$, $k_0 = 1$, $l_B = 1$ and this information is revealed to all nodes.

*Decoding:* Destination 1 receives $y_{1,b}^n$ at the end of block $b$. We assume the decoding is correct in previous blocks. The decoder first tries to find an index $\hat{k}_{b-1}$ such that $(x_R^n(\hat{k}_{b-1}), y_{1,b}^n) \in T_\epsilon^n$. It then searches for $\hat{w}_{1,b-1}$ such that $(x_1^n(\hat{w}_{1,b-1}), x_2^n(\hat{w}_{2,b-1}), x_R^n(\hat{k}_{b-2}), \hat{y}_R^n(\hat{l}_{b-1}|\hat{k}_{b-2}), y_{1,b-1}^n) \in T_\epsilon^n$ for some $\hat{w}_{2,b-1}$ and $\hat{l}_{b-1} \in \mathcal{B}(\hat{k}_{b-1})$. Destination 2 follows similar decoding steps.

*Error Analysis:* Assume $w_{1,b-1} = 1, w_{2,b-1} = 1, k_{b-1} = K_{b-1}, k_{b-2} = K_{b-2}, l_{b-1} = L_{b-1}$. Define the following error events.

$$\mathcal{E}_0 := \{(Y_{R,b}^n, \hat{Y}_R^n(l_b|K_{b-1}), X_R^n(K_{b-1})) \notin T_\epsilon^n, \forall l_b\}$$

$$\mathcal{E}_{10} := \{(X_R^n(K_{b-1}), Y_{1,b}^n) \notin T_\epsilon^n\}$$

$$\mathcal{E}_{11} := \{\exists \hat{k}_{b-1} \neq K_{b-1} : (X_R^n(\hat{k}_{b-1}), Y_{1,b}^n) \in T_\epsilon^n \text{ for all } \hat{k}_{b-1}\}$$

$$\mathcal{E}_{12} := \{(X_1^n(1), X_2^n(1), X_R^n(\hat{K}_{b-2}), \hat{Y}_R^n(L_{b-1}|\hat{K}_{b-2}), Y_{1,b-1}^n) \notin \mathcal{T}_\epsilon^n\}$$

$$\mathcal{E}_{13} := \{\exists \hat{w}_{1,b-1} \neq 1 : (X_1^n(\hat{w}_{1,b-1}), X_2^n(1), X_R^n(\hat{K}_{b-2}), \hat{Y}_R^n(L_{b-1}|\hat{K}_{b-2}), Y_{1,b-1}^n) \in \mathcal{T}_\epsilon^n\}$$

$$\mathcal{E}_{14} := \{\exists \hat{w}_{1,b-1} \neq 1 : (X_1^n(\hat{w}_{1,b-1}), X_2^n(\hat{w}_{2,b-1}), X_R^n(\hat{K}_{b-2}), \hat{Y}_R^n(L_{b-1}|\hat{K}_{b-2}), Y_{1,b-1}^n) \in \mathcal{T}_\epsilon^n \text{ for some } \hat{w}_{2,b-1} \neq 1\}$$

$$\mathcal{E}_{15} := \{\exists \hat{w}_{1,b-1} \neq 1 : (X_1^n(\hat{w}_{1,b-1}), X_2^n(1), X_R^n(\hat{K}_{b-2}), \hat{Y}_R^n(\hat{l}_{b-1}|\hat{K}_{b-2}), Y_{1,b-1}^n) \in \mathcal{T}_\epsilon^n$$
$$\text{for some } \hat{l}_{b-1} \in \mathcal{B}(\hat{K}_{b-1}), \hat{l}_{b-1} \neq L_{b-1}\}$$
$$\mathcal{E}_{16} := \{\exists \hat{w}_{1,b-1} \neq 1 : (X_1^n(\hat{w}_{1,b-1}), X_2^n(\hat{w}_{2,b-1}), X_R^n(\hat{K}_{b-2}), \hat{Y}_R^n(\hat{l}_{b-1}|\hat{K}_{b-2}), Y_{1,b-1}^n) \in \mathcal{T}_\epsilon^n$$
$$\text{for some } \hat{l}_{b-1} \in \mathcal{B}(\hat{K}_{b-1}), \hat{l}_{b-1} \neq L_{b-1}, \hat{w}_{2,b-1} \neq 1\}$$

$\mathcal{E}_{20} - \mathcal{E}_{26}$ are defined in similar fashion by switching indices 1 and 2. The error probability is

$$P(\mathcal{E}) = P\Big(\mathcal{E}_0 \bigcup \cup_{i=0}^6 \mathcal{E}_{1i} \bigcup \cup_{i=0}^6 \mathcal{E}_{2i}\Big) \tag{66}$$
$$\leq P(\mathcal{E}_0) + P(\mathcal{E}_{10}) + P(\mathcal{E}_{11}) + P(\mathcal{E}_{20}) + P(\mathcal{E}_{21}) + \sum_{i=3}^{6} P\Big(\mathcal{E}_{1i} \bigcap (\mathcal{E}_0^c \cap \mathcal{E}_{10}^c \cap \mathcal{E}_{11}^c \cap \mathcal{E}_{20}^c \cap \mathcal{E}_{21})\Big)$$
$$+ \sum_{i=3}^{6} P\Big(\mathcal{E}_{2i} \bigcap (\mathcal{E}_0^c \cap \mathcal{E}_{10}^c \cap \mathcal{E}_{11}^c \cap \mathcal{E}_{20}^c \cap \mathcal{E}_{21})\Big)$$

According to covering lemma [13], $P(\mathcal{E}_0) \to 0$ as $n \to \infty$ as long as

$$\hat{R} > I(Y_R; \hat{Y}_R | X_R) \tag{67}$$

According to Law of Large Numbers (LLN), $P(\mathcal{E}_{10}) \to 0$ and $P(\mathcal{E}_{20}) \to 0$ as $n \to \infty$. According to packing lemma [13], $P(\mathcal{E}_{11}) \to 0$ and $P(\mathcal{E}_{21}) \to 0$ as $n \to \infty$ as long as

$$R_0 < \min\{I(X_R; Y_1), I(X_R; Y_2)\} \tag{68}$$

According to conditional typicality lemma [13], $P(\mathcal{E}_{12}) \to 0$ and $P(\mathcal{E}_{22}) \to 0$ as $n \to \infty$. Following the derivation in [13], we can show that $P(\mathcal{E}_{13}) \to 0$, $P(\mathcal{E}_{14}) \to 0$, $P(\mathcal{E}_{15}) \to 0$, $P(\mathcal{E}_{16}) \to 0$ as $n \to \infty$ as long as

$$R_1 < I(X_1; \hat{Y}_R Y_1 | X_2 X_R) \tag{69}$$
$$R_1 + R_2 < I(X_1 X_2; \hat{Y}_R Y_1 | X_2 X_R) \tag{70}$$
$$R_1 + \hat{R} - R_0 < I(X_1; Y_1 | X_2 X_R) + I(\hat{Y}_R; X_1 X_2 Y_1 | X_R) \tag{71}$$
$$R_1 + R_2 + \hat{R} - R_0 < I(X_1 X_2; Y_1 | X_R) + I(\hat{Y}_R; X_1 X_2 Y_1 | X_R) \tag{72}$$

The analysis for events $\mathcal{E}_{23}, \mathcal{E}_{24}, \mathcal{E}_{25}, \mathcal{E}_{26}$ follows similar steps. Combining these constraints with

(67) and (68), we obtain the rate constraints in $GCF_2$. ∎

# Appendix C

# Proof of Lemma 1

*Proof:* We first rewrite (39) and (40) as

$$h(h_{12}X_1^n + Z_2^n, h_{1R}X_1^n + Z_R^n | N_2^n, N_4^n) = h(h_{12}X_1^n + W_2^n, h_{1R}X_1^n + W_4^n)$$

$$h(h_{21}X_2^n + Z_1^n, h_{2R}X_2^n + Z_R^n | N_1^n, N_3^n) = h(h_{21}X_2^n + W_1^n, h_{2R}X_2^n + W_3^n)$$

where $W_i \sim \mathcal{N}(0, 1-\rho_i^2),\ i = 1,2,3,4$. $W_1, W_2, W_3, W_4$ are independent of $N_1, N_2, N_3, N_4$, respectively.

$$\begin{aligned}
& h(h_{12}X_1^n + h_{12}N_1^n, h_{1R}X_1^n + h_{1R}N_3^n) - h(h_{12}X_1^n + W_2^n, h_{1R}X_1^n + W_4^n) \\
& = h(h_{1R}X_1^n + h_{1R}N_3^n) + h(h_{12}X_1^n + h_{12}N_1^n | h_{1R}X_1^n + h_{1R}N_3^n) \\
& \quad - h(h_{1R}X_1^n + W_4^n) - h(h_{12}X_1^n + W_2^n | h_{1R}X_1^n + W_4^n) \qquad (73) \\
& = h(h_{1R}X_1^n + h_{1R}N_3^n) + h(h_{12}N_1^n - h_{12}N_3^n | h_{1R}X_1^n + h_{1R}N_3^n) \\
& \quad - h(h_{1R}X_1^n + W_4^n) - h(W_2^n - \frac{h_{12}}{h_{1R}}W_4^n | h_{1R}X_1^n + W_4^n) \qquad (74) \\
& = h(h_{1R}X_1^n + h_{1R}N_3^n | h_{12}N_1^n - h_{12}N_3^n) + h(h_{12}N_1^n - h_{12}N_3^n) \\
& \quad - h(h_{1R}X_1^n + W_4^n | W_2^n - \frac{h_{12}}{h_{1R}}W_4^n) - h(W_2^n - \frac{h_{12}}{h_{1R}}W_4^n) \\
& = h(h_{1R}X_1^n + V_{13}^n) + h(h_{12}N_1^n - h_{12}N_3^n) - h(h_{1R}X_1^n + U_{24}^n) - h(W_2^n - \frac{h_{12}}{h_{1R}}W_4^n) \qquad (75)
\end{aligned}$$

where

$$V_{13} \sim \mathcal{N}(0, \frac{h_{1R}^2\sigma_1^2\sigma_3^2}{\sigma_1^2 + \sigma_3^2}),\ U_{24} \sim \mathcal{N}(0, \frac{h_{1R}^2(1-\rho_2^2)(1-\rho_4^2)}{h_{1R}^2(1-\rho_2^2) + h_{12}^2(1-\rho_4^2)})$$

From the worst case noise lemma in [28], we have

$$h(h_{1R}X_1^n + V_{13}^n) - h(h_{1R}X_1^n + U_{24}^n) \leq$$

$$nh(h_{1R}X_{1G} + V_{13}) - nh(h_{1R}X_{1G} + U_{24})$$

if

$$\frac{h_{1R}^2\sigma_1^2\sigma_3^2}{\sigma_1^2+\sigma_3^2} \leq \frac{h_{1R}^2(1-\rho_2^2)(1-\rho_4^2)}{h_{1R}^2(1-\rho_2^2)+h_{12}^2(1-\rho_4^2)}$$

where $X_{1G} \sim \mathcal{N}(0, P_1)$, which gives us the condition (41). Using similar method we can obtain the condition (42). ■

## APPENDIX D
## PROOF OF LEMMA 2

*Proof:*

$$I(X_{1G}; S_1, S_R|Y_1, Y_R) \tag{76}$$

$$= h(S_1, S_R|Y_1, Y_R) - h(S_1, S_R|Y_1, Y_R, X_{1G}) \tag{77}$$

$$= h(S_1|Y_1, Y_R) + h(S_R|S_1, Y_1, Y_R) - h(S_1|Y_1, Y_R, X_{1G}) - h(S_R|S_1, Y_1, Y_R, X_{1G}) \tag{78}$$

$$\leq h(S_1|Y_1) + h(S_R|Y_R) - h(S_1|Y_1, Y_R, X_{1G}) - h(S_R|S_1, Y_1, Y_R, X_{1G}) \tag{79}$$

$$= h(X_{1G} + N_1|X_{1G} + h_{21}X_{2G} + Z_1) - h(N_1|h_{21}X_{2G} + Z_1)$$

$$+ h(h_{1R}X_{1G} + h_{1R}N_3|h_{1R}X_{1G} + h_{2R}X_{2G} + Z_R) - h(h_{1R}N_3|h_{2R}X_{2G} + Z_R) \tag{80}$$

As long as (80) is 0, the genie $S_1, S_R$ is smart. We can perform similar operation for the other term $I(X_{2G}; S_2, T_R|Y_2, Y_R)$, and conditions (44) can be obtained. ■